\documentclass[%
reprint,
superscriptaddress,
amsmath,amssymb,
aps,
pra, nh
]{revtex4-2}
\usepackage{graphicx}
\usepackage{dcolumn}
\usepackage{bm}
\usepackage{hyperref}
\usepackage{braket}
\usepackage{multirow}

\usepackage{xcolor}
\usepackage[normalem]{ulem}

\newcommand{\blue}[1]{\textcolor{black}{#1}}

\newcommand{\expval}[1]{\left<#1\right>}

\newcommand{\piflux}{$\pi$\textrm{-flux }}
\begin{document}
	\title{Confined states and topological phases in two-dimensional quasicrystalline \piflux model}
	\author{Rasoul Ghadimi*}
	\affiliation{Center for Correlated Electron Systems, Institute for Basic Science (IBS), Seoul 08826, Korea}
	\affiliation{Department of Physics and Astronomy, Seoul National University, Seoul 08826, Korea}
	\affiliation{Center for Theoretical Physics (CTP), Seoul National University, Seoul 08826, Korea}

	\author{Masahiro Hori*}
	\affiliation{Department of Applied Physics, Tokyo University of Science, Tokyo 125-8585, Japan}
	\thanks{\blue{These authors contributed equally}}
 
	\author{Takanori Sugimoto}
	\affiliation{Center for Qunatum Information and Qunatum Biology, Osaka University, Osaka 560-0043, Japan}
	\affiliation{Advanced Science Research Center, Japan Atomic Energy Agency, Tokai, Ibaraki 319-1195, Japan}
	
	\author{Takami Tohyama}
	\affiliation{Department of Applied Physics, Tokyo University of Science, Tokyo 125-8585, Japan}
	\date{\today}
	
	\begin{abstract}
		Motivated by topological equivalence between an extended Haldane model and a chiral-\piflux model on a square lattice, we apply \piflux models to two-dimensional bipartite quasicrystals with rhombus tiles in order to investigate topological properties in aperiodic systems.
		Topologically trivial \piflux models in the Ammann-Beenker tiling lead to massively degenerate confined states whose energies and fractions differ from the zero-flux model. This is different from the \piflux models in the Penrose tiling, where confined states only appear at the center of the bands as is the case of a zero-flux model.
		Additionally, Dirac cones appear in a certain \piflux model of the Ammann-Beenker approximant, which remains even if the size of the approximant increases.
		Nontrivial topological states with nonzero Bott index are found when staggered tile-dependent hoppings are introduced in the \piflux models. 
		This finding suggests a new direction in realizing nontrivial topological states without a uniform magnetic field in aperiodic systems.
	\end{abstract}
	\maketitle
	\section{Introduction}
	The discovery of exotic phases in quasicrystals, such as quantum criticality~\cite{Deguchi2012}, superconductivity~\cite{Kamiya2018}, and magnetism~\cite{Tamura2021}, has stimulated further research about the role of aperiodicity in establishing these phenomena. Although quasicrystals lack translational symmetry~\cite{PhysRevLett.53.1951,PhysRevB.34.596,PhysRevB.34.617}, they possess unique structural properties such as self-similarity, originating from their higher-dimensional structures. Recent advances in realizing quasicrystals enable researchers to analyze aperiodicity in combination with other physical phenomena~\cite{PhysRevLett.79.3363,PhysRevX.6.011016,Layer_dependent_topological_quasicrystals_Jeffrey2020,PhysRevLett.122.110404}.
	Further efforts are needed to understand and explore classical and quantum phases in such systems~\cite{PhysRevB.43.1378,PhysRevB.35.1456,PhysRevB.38.5981,PhysRevB.37.2797,PhysRevB.38.1621,PhysRevLett.66.333,Watanabe2021}\blue{\cite{PhysRevLett.126.110401,PhysRevA.72.053607}} including topological phases of matter, which have become a central issue in various fields of condensed matter physics~\cite{RevModPhys.83.1057,doi:10.1146/annurev-conmatphys-062910-140432,RevModPhys.82.3045}.
	Interestingly, aperiodicity in quasicrystals does not prevent them from hosting topological phases~\cite{PhysRevLett.111.226401,PhysRevLett.109.106402,PhysRevB.100.165101,PhysRevLett.119.215304,PhysRevLett.110.076403,PhysRevX.9.021054,PhysRevLett.123.196401,PhysRevResearch.2.033071,doi:10.7566/JPSJ.86.114707,Quasicrystalline_Weyl_2022_Fonseca,PhysRevB.103.085307,PhysRevB.104.155304,PhysRevB.100.085119,PhysRevB.101.041103,PhysRevB.98.165427}\blue{\cite{PhysRevB.106.045417,PhysRevB.104.245302}} including Chern insulators~\cite{PhysRevB.94.205437,PhysRevLett.121.126401,PhysRevB.98.125130,Topological_quasicrystals_Fan2021,PhysRevB.101.115413,PhysRevB.106.L201113,PhysRevB.100.214109,PhysRevB.100.115311,PhysRevB.91.085125}, topological superconductors~\cite{PhysRevLett.116.257002,PhysRevLett.125.017002,PhysRevB.104.144511,PhysRevB.95.024509,PhysRevB.100.014510}, non-Hermitian topological phases~\cite{PhysRevLett.122.237601,PhysRevB.105.014202}, and higher-order topological phases~\cite{PhysRevLett.124.036803,PhysRevB.102.241102,PhysRevLett.129.056403,Lv2021_quasicrystalline_quadrupole}.
	
	\begin{figure*}
		\includegraphics[width=0.8\textwidth]{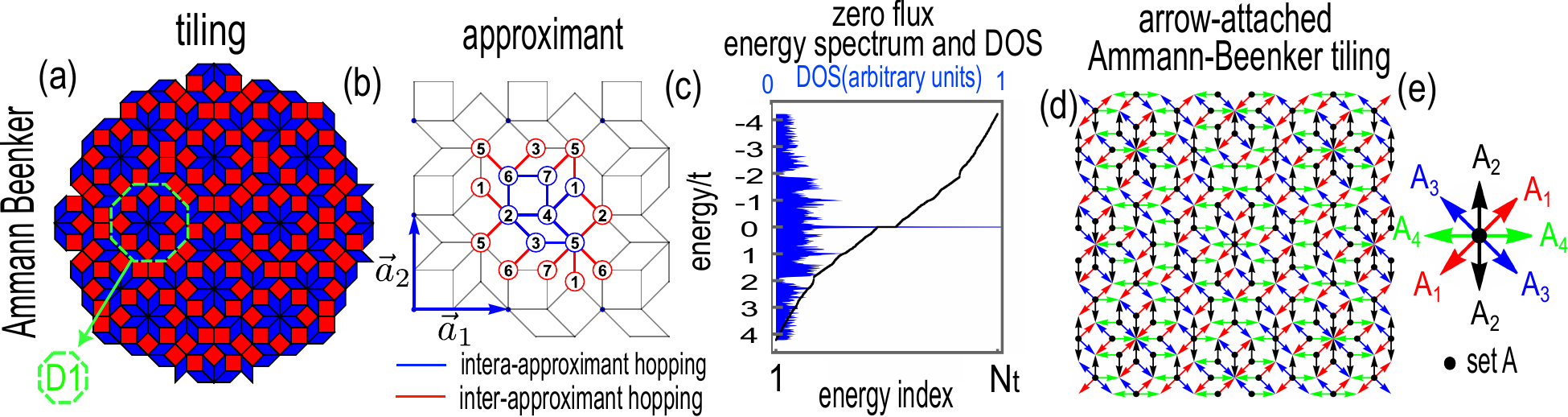}
		\caption[quasicrystal]{
			(a) Ammann-Beenker tiling.
                The indicated area with green dashed edges refers to a symmetric domain of tiles called $D_1$ that is used in TABLE.~\ref{table:conf1E1NiNet}. 
			(b) An example of Ammann-Beenker tiling approximant ($g=1$).
			The numbers inside circles show approximant  sublattice indices. Blue dots show the origin of a unit-cell and blue arrows represent the basis vectors of the unit-cell, $\vec{a}_1$, $\vec{a}_2$.
			(c)  Energy levels (black points) sorted from lowest to highest energies versus index of levels, 1 to $N_t$ in the bottom horizontal axis and DOS (blue lines) for a zero-flux model in an Ammann-Beenker approximant [$N_t=2\times2\times1393$]. 
			(d) Edges of the Ammann-Beenker tiling drawn by arrows starting from a vertex of the set A.
			The colored arrows correspond to those in (e). (e) Phase $A_n$ ($=\pi/2$, $n=1,\dots,4$) to construct  quasicrystalline \piflux models. The direction of arrows corresponds to $\hat{R}_n$ (see the text).}
		\label{fig:tiling}
	\end{figure*}
	
	In this \blue{paper}, we present a new proposal for generalizing the Haldane model~\cite{PhysRevLett.61.2015} on a honeycomb lattice to quasicrystalline systems.
	The Haldane model describes nontrivial topological states without an external magnetic field.
	While previous studies have generalized this model to quasicrystalline systems~\cite{PhysRevB.100.214109,PhysRevB.106.L201113}, a procedure similar to a conventional generalization using \piflux models~\cite{PhysRevLett.106.236804,PhysRevB.39.11413,PhysRevB.74.205414,Wang_2012,Li_2008} applied to other translational systems such as a square lattice has not been employed.
	We extend the \piflux model to two-dimensional bipartite quasicrystals with rhombus tiles, where $\pi$ flux penetrates tiles with specific configurations.
	As examples of such quasicrystalline systems, we analyze two quasicrystals, Penrose tiling, and Amman-Beenker tiling ~\cite{PhysRevB.39.10519}.
	We find that in contrast to zero flux limit, \piflux configurations in the Amman-Beenker tiling introduce massively degenerate confined states whose energies are not at the center of energy bands. These states are strictly localized, similar to well-known confined states~\cite{PhysRevLett.55.2915,PhysRevLett.56.2740,PhysRevB.38.12903,PhysRevB.51.15827,PhysRevB.106.064207,PhysRevB.96.214402,PhysRevB.106.024201,PhysRevB.102.115125,PhysRevB.104.014204,PhysRevB.102.064213,PhysRevB.104.165112,matsubara2022confined} and dispersionless flat bands of periodic kagome and Lieb lattices, etc.~\cite{PhysRevLett.62.1201,PhysRevB.34.5208,PhysRevB.54.R17296,PhysRevB.99.045107,Dias2015,doi:10.1080/23746149.2021.1901606,PhysRevB.104.L081104,doi:10.1080/23746149.2018.1473052,PhysRevLett.71.4389,Hwang2021,Rhim2020}.
	The massive degeneracy of the confined states is related to the self-similarity of quasicrystals, where local patterns repeat through quasicrystals due to Conway’s theorem~\cite{PhysRevB.96.214402}. 
	Accordingly, if a localized wave function appears in a small part of a given quasicrystal, it also appears in the repeated regions, leading to massively degenerate confined states.
	On the other hand, in the Penrose tiling, confined states always appear at the center of energy bands regardless of the presence of \piflux, and their number remains constant despite modifications to their wave function induced by the \piflux model.
        This is related to the local violation of bipartite neutrality in the Penrose tiling~\cite{PhysRevB.102.064210}.
	We investigate the confined states  by developing a method to obtain their localized wave function and analyzing their fraction in different \piflux configurations.
	Furthermore, we discover Dirac cones in the energy dispersion of a particular \piflux configuration in Ammann-Beenker approximants.
	We find nontrivial topological states in all of the \piflux configurations in both Penrose tiling and Amman-Beenker tiling when we introduce staggered tile-diagonal hopping (STDH). 
	We confirm the emergence of topological states in the \piflux models with STDH by calculating the Bott index and edge modes.
	Therefore, this model can be a new proposal for the generalization of the Haldane model to aperiodic systems in realizing nontrivial topological states without a uniform magnetic field. 
 
	\blue{The structure of this paper is as follows. In Sec.\ref{SEC:AB}, we focus on the Ammann-Beenker tiling, while in Sec.\ref{SEC:PEN}, we explore the Penrose tiling. Finally, our findings are summarized and discussed in Sec.~\ref{SEC:SUM}.}

        \section{Ammann-Beenker \piflux model}\label{SEC:AB}
	\textit{Structure--} Ammann-Beenker tiling consists of two different tiles (see Fig.~\ref{fig:tiling}(a)) \cite{PhysRevB.42.8091}. 
	We regard vertices of the tiles as positions where electrons can hop through the edges of tiles. 
	This is known as the vertex model~\cite{PhysRevB.58.13482,PhysRevB.38.1621}.
	This model can be studied by using Ammann-Beenker approximants, or equivalently supercell (see Fig.~\ref{fig:tiling}(b))~\cite{PhysRevB.43.8879,PhysRevB.102.224201,Duneau_1989}, where periodic boundary conditions are employed at the edges of the supercell. 
        \blue{Approximants are standard tools to investigate topological phases  in quasicrystals \cite{Topological_quasicrystals_Fan2021}.}
        The size of quasicrystal approximants can be increased by increasing the approximant generation $g$.
	By definition, an approximant supercell contains full information of a given quasicrystal when $g\rightarrow\infty$.
        \blue{Intuitively, approximant is a defective quasicrystal, where in some tiles matching rule of the given tiling is broken down. However, the number of these defects decreases in comparison to the number of all tiles by increasing approximant generation \cite{PhysRevB.43.8879}.  }
	In order to reduce boundary effects and to perform Bott index calculations, we use approximants with $g=4$, which gives $N_t=1393$ total vertices.

	The vertex model for the Ammann-Beenker tiling has bipartite properties. 
	This means that we can decompose vertices of the Ammann-Beenker tiling into two sets, A and B, without any edges connecting the same set~\cite{PhysRevB.96.161104}.
	As a result, starting from a given vertex, we cannot make a loop with an odd number of edges. 
	However, our method (see Ref.~\onlinecite{Duneau_1989}) to generate the approximant of the Ammann-Beenker tiling violates bipartite properties at the boundaries (e.g. a \textcircled{1} $\rightarrow$ \textcircled{2} $\rightarrow$ \textcircled{4} $\rightarrow$ \textcircled{1} path in Fig.~\ref{fig:tiling}(b)). 
        \blue{Note that,  to calculate the Bott index we need to have a correct periodic boundary condition (which respects the \piflux model that defines shortly), which is violated if we use a single cell of approximant of Ammann-Beenker tiling.}
	Nonetheless, we can restore bipartite properties by constructing a new approximant made from $2\times 2$ non-bipartite approximants. 
	In the following, we will use this new approximant to study the Ammann-Beenker tiling.

        \textit{General Hamiltonian--} Hamiltonian for electrons in Ammann-Beenker tiling with arbitrary flux distribution reads
        \begin{equation}\label{eq:APM}
		H=\sum_{\expval{i,j}} t e^{i A_{ij}} \ket{i}\bra{j},
	\end{equation}
	where $t$ and $A_{ij}=-A_{ji}$ are hopping amplitude and Peierls  phase, respectively. 
	 $\expval{i,j}$ indicates all edges of the Ammann-Beenker tiling~\cite{PhysRevB.106.L201113} and $A_{ij}$ gives a complex phase for the hopping.
	The total flux penetrating each face or tile is given by the total phases that an electron can accumulate along edges, i.e.,
        \begin{equation}\label{key}
	\phi_{i,j,k,l}=A_{ij}+A_{jk}+A_{kl}+A_{li},
	\end{equation}
	where $i$, $,j$, $k$, and $l$ represent vertices on a given tile and are arranged in an anticlockwise way such as $i\rightarrow j\rightarrow k \rightarrow l \rightarrow i$.

	\textit{Zero flux case--} 
        Let us first review the zero flux limit, where we set $A_{ij}=0$ in Eq.~(\ref{eq:APM}).
 In Fig.~\ref{fig:tiling}(c), we show the energy levels and density of states (DOS) of the zero-flux Ammann-Beenker approximant with $N_t=2\times2\times1393$.
	Both energy levels and DOS are symmetric with respect to the energy $E=0$, reflecting bipartite nature. 
	A prominent peak exists in DOS at $E=0$, which corresponds to the presence of confined states~\cite{PhysRevB.102.115125}.
	In the Ammann-Beenker tiling, the fraction of confined states $p$ (the number of them divided by $N_t$) is given by $p=p_{E=0}=1/2\tau_s^2\approx0.086$~\cite{PhysRevB.102.115125}, where $\tau_s=1+\sqrt{2}$ is the silver ratio.
	Since the confined states in the Ammann-Beenker tiling are fragile, they generally disperse upon introducing other terms into the Hamiltonian~\cite{PhysRevB.102.115125}. 
	Furthermore, the confined states are mixed with other states. This may be related to the fact that increasing the system size generates a new confined state with larger extension~\cite{PhysRevB.102.115125}.

	\begin{figure*}[t!]
		\centering
		\includegraphics[width=0.95\linewidth]{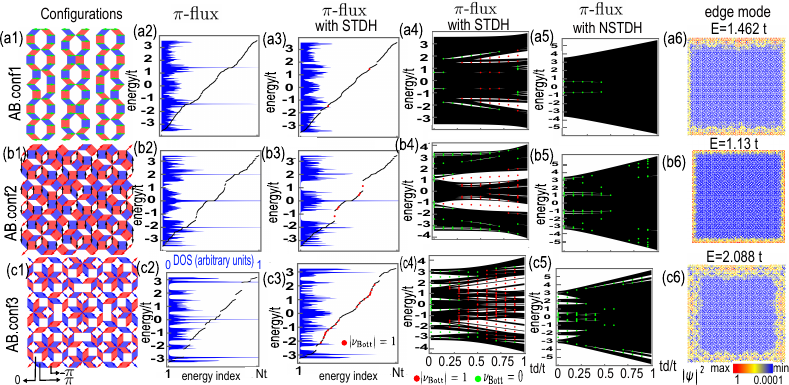}
		\caption[Results]{(a1), (b1), and (c1): different \piflux configurations named AB.conf1-3 for the Ammann-Beenker tiling, where red, blue, and white regions indicate tiles with $-\pi$, $+\pi$, and zero flux, respectively.
			Arrows indicate the presence of nonzero phases on the edge of tiles. 
			Energy levels and DOS for (a2) AB.conf1, (b2) AB.conf2, and (c2) AB.conf3.
			(a3), (b3), (c3): same as (a2), (b2), (c2) but STDH with $t_d=0.3t$ is used.
			The energy levels as a function of STDH $t_d$ for (a4) AB.conf1, (b4) AB.conf2, and (c4) AB.conf3.
			(a5), (b5), (c5): same as (a4), (b4), (c4) but NSTDH is used.
			Edge modes at a given energy $E$ in the case of STDH $t_d=0.3t$ for (a6) AB.conf1, (b6) AB.conf2, and (c6) AB.conf3, where the red (blue) region indicates a larger (smaller) probability of an edge-mode wave function.}
		\label{fig:resultAB}
	\end{figure*}

	\textit{Quasicrystalline \piflux model--}
	Following the \piflux model on a square lattice, which introduces \piflux systematically~\cite{PhysRevLett.106.236804} (see Appendix~\ref{sec:pifluxsquare}), we assume the same phase $A_{ij}$ for edges with the same orientation, where the edges start from a vertex of the set A.
	However, for the edges starting from a vertex of set B, we use the same phase but with the opposite sign to keep the hermiticity of the Hamiltonian.
	Therefore, it is useful to define phase $A_n$ for the edges starting from a vertex of the set A with a directional vector $\hat{R}_n=\pm(\cos(2\pi n/8), \sin(2\pi n/8))$, where $n=1, \dots,4$.
        The quasicrystalline \piflux model is defined by setting some (but not all) of $A_n$ to $\pm\pi/2$ and others to zeros.
        Note that as fluxes $\{\phi_{i,j,k,l}\}$ are defined modules to $2\pi$, setting all $A_n$ to $\pm\pi/2$  results in a zero flux model.
        
	In Fig.~\ref{fig:tiling}(d), we present the Ammann-Beenker tiling with attached arrows that are defined based on $A_n$ (see Fig.~\ref{fig:tiling}(e)).
	We note that one can construct this arrow-attached tiling using inflation rules (see Appendix~\ref{sec:inflation}). 
	Furthermore, it is worth mentioning that the arrows with the same colors construct structures similar to Ammann lines in the Ammann-Beenker tiling (for example, see green arrows in Fig.~\ref{fig:resultAB} (a1))~\cite{PhysRevB.39.10519}.

	\textit{Confined states in Ammann-Beenker \piflux model--}
	The Ammann-Beenker tiling allows various \piflux configurations due to possible choices of nonzero $A_n$.
	In the first column of Fig.~\ref{fig:resultAB}, we show three possible configurations of \piflux in the Ammann-Beenker tiles.
	In the following, we name the three cases AB.conf1, AB.conf2, and AB.conf3, as shown in Figs.~\ref{fig:resultAB}(a1),  \ref{fig:resultAB}(a2), and \ref{fig:resultAB}(a3), respectively.
	\blue{We note that these configurations are unique and  other configurations are given by rotation and/or translation of the three configurations}. 
	The AB.conf1 is obtained by choosing $A_4=\pi/2$~and~$A_{n\ne4}=0$; AB.conf2 is by $A_1=A_2=\pi/2$~and~$A_{n\ne1,2}=0$; and AB.conf3 is by $A_1=A_3=\pi/2$~and~$A_{n\ne1,3}=0$.
	These \piflux configurations lead to confined states with different energies and fraction.
	We plot energy levels and DOS for AB.conf1, AB.conf2, and AB.conf3, in Figs.~\ref{fig:resultAB}(a2), \ref{fig:resultAB}(b2), and \ref{fig:resultAB}(c2), respectively. Confined states appear at $E=\pm \sqrt{2}t$ for AB.conf1, at $E=0$ and $\pm 2t$ for AB.conf2, and $E=0$ and $\pm2\sqrt{2}t$ for AB.conf3.
	
We can confirm that confined states actually consist of localized wave functions after some manipulations of degenerate eigenfunctions numerically obtained by exact diagonalization. 
	Usually such eigenfunctions, i.e.,  wave functions, give finite overlap at the same vertex.
	This means that they are not localized wave functions.
	However,  one can construct localized wave functions by recombining the degenerate wave functions.
	For this purpose, we define inverse participant ratio (IPR) given by $\sum_i |\Psi_i|^4/\sum_i |\Psi_i|^2$, where $\Psi_i$ is the $i$-th vertex component of a variational wave function composed of linearly combined degenerate wave functions.
	IPR gives a measure of the extension of a given wave function.
	For instance, IPR for extended  Bloch states in translationally invariant systems is zero, while it gives $1$ for atomic wave-functions~\cite{PhysRevB.105.045146,PhysRevB.100.014510}.
	Maximizing IPR, we can obtain localized wave functions corresponding to confined states (see Appendix~\ref{sec:IPR}). 
	After confirming a localized wave function for each confined state, we perform an analysis similar to Ref.~\cite{PhysRevB.102.115125} to obtain the fraction $p$ of the confined states.
 \blue{}
	We conjecture $p$ for each configuration as $p=p^{\text{AB.conf1}}_{E=\pm\sqrt{2}t}=\tau_{s}^{-4}\approx0.029$, $p=p^{\text{AB.conf2}}_{E=0}\approx0.061$, $p=p^{\text{AB.conf2}}_{E=\pm2t}=198-82\tau_s\approx0.034$, $p^{\text{AB.conf3}}_{E=0}=2675 - 1108 \tau_s\approx 0.051$, and $p=p^{\text{AB.conf3}}_{E=\pm2\sqrt{2}t}=\tau_{s}^{-4}\approx0.029$ (see TABLE.~\ref{table:conf1E1NiNet}, and Appendix~\ref{sec:confineAmmann} for further analysis).
 \blue{The investigation of the fraction of confined states involves analyzing their occurrence in smaller structures and extrapolating their statistics to larger systems. It is important to acknowledge that there is no guarantee of the emergence of new confined states in larger system sizes, which leads us to employ the term "conjecture". 
 In particular, for the $E=0$ confine states on AB.Conf2, we were unable to find a suitable expression for precisely determining the fraction of confined states.}

 \begin{table}[t!]
		\caption{The number of newly generated confined states $N_{j}^{\rm{net}}$ in the $D_j$ region for the Ammann-Beenker tiling, where $D_j$ defines a cluster of tiles that appears after $(j-1)$-th inflation of $D_1$ indicated by green in Fig.~\ref{fig:tiling}(a). 
	$N_{j}^{\rm{net}}$ is calculated for $j=1,\dots,5$ and presented in each column for confined states with energy $E$ in a given flux configuration.
For $j>5$, we use speculated $N_{j}^{\rm{net}}$ expected from the calculated $N_{j}^{\rm{net}}$ for $j=1,\dots,5$ (see Ref.~\onlinecite{PhysRevB.102.115125} and Appendix~\ref{sec:confineAmmann} for further analysis). 
	The ? mark means that it is difficult to speculate the number.
	The last column represents the fraction of confined states defined by $p=\sum_j p_j N_j^{\text{net}}$, where $p_j=2\tau_s^{-(2j+3)}$ is the fraction for the $D_j$ region.
  The fraction of confined states $p$ for AB.conf2 and $E=0$, is calculated numerically for a Ammann-Beenker tiling with $N_t\approx 2\times10^4$ vertices. }
		\label{table:conf1E1NiNet}
  		\resizebox{\columnwidth}{!}{	
    \begin{tabular}{c||c c c c c c c}
     
			&$N_{1}^{\rm{net}}$&$N_{2}^{\rm{net}}$&$N_{3}^{\rm{net}}$&$N_{4}^{\rm{net}}$&$N_{5}^{\rm{net}}$&$N_{i}^{\rm{net}}$&$p$\\
				\hline\hline
     $\begin{array}{c}
       \text{zero flux}     \\
         E=0
     \end{array}$&2&6&12&20&30& $i(i+1)$&$1/2\tau_s^2$\\\hline
    $\begin{array}{c}
       \text{AB.conf1}     \\
         E=\pm\sqrt2t
     \end{array}$  & 1&1&1&1&1&1&$1/\tau_s^4$\\\hline
    $\begin{array}{c}
       \text{AB.conf2}     \\
         E=0
     \end{array} $&1&4&15&68&347&?&$\approx0.061$\\\hline
    $\begin{array}{c}
       \text{AB.conf2 }     \\
         E=0, t_d\ne0
     \end{array}$&1&2&3&4&5&i&$1/2\tau_s^3$ \\\hline
    $\begin{array}{c}
       \text{AB.conf2}     \\
         E=\pm2t
     \end{array}$& 1&2&2&2&2&2&\scriptsize{$198{-}82\tau_s$}\\\hline
    $\begin{array}{c}
       \text{AB.conf3}     \\
         E=0
     \end{array}$& 1&5&7&7&7&7&\scriptsize{$2675{-}1108 \tau_s$}\\\hline
    $\begin{array}{c}
       \text{AB.conf3}     \\
         E=\pm2\sqrt2t
     \end{array}$& 1&1&1&1&1&1&$1/\tau_s^4$\\
		\end{tabular}
  }
	\end{table}

	\textit{Topological phase--}
	It has been known that topological states emerge in a periodic \piflux model on a square lattice if STDH is set to be $\pm t_d$ for tiles with $\pm \pi$ fluxes~\cite{PhysRevLett.106.236804} (see Appendix~\ref{sec:pifluxsquare}). 
        We introduce the same STDH in our Ammann-Beenker \piflux configurations.
	However, in our Ammann-Beenker \piflux model, we obtain some tiles without any flux and we assume no diagonal hopping for them.
        \blue{We note that, prior to considering the STDH, the fact that all fluxes exhibit a \piflux distribution, indicates the presence of time-reversal symmetry. This can be seen as follows: consider, for example, the distribution shown in Fig.~\ref{fig:resultAB}(a1). By choosing alternative hoppings, such as $\pm1$ for the arrows along the strips, the same \piflux distribution can be achieved, further demonstrating the presence of time-reversal symmetry. However, with the introduction of the STDH, the tiles become half and acquire a $\pi/2$ flux distribution, leading to the breaking of time-reversal symmetry.  }
	We plot energy levels and DOS for $t_d=0.3 t$ in the third column of Fig.~\ref{fig:resultAB} for the three \piflux configurations.
	In all cases except AB.conf2, introducing STDH disperses the confined states for $t_d=0$.
        We found that in AB.conf2 actually STDH disperse approximately half of the confine states for $t_d=0$ and their fraction is approximately given by $p=p^{\text{AB.conf2}}_{E=0,t_d\ne0}=1/2\tau_s^3\approx0.035$ (see  TABLE.~\ref{table:conf1E1NiNet}).
	Additionally, nonzero $t_d$ develops new gaps in their energy-level spectrum. 
	We calculate the Bott index $\nu_{\text{Bott}}$~\cite{Toniolo2022,HASTINGS20111699,Loring2019AGT} (see Appendix~\ref{sec:Bott}), which is equivalent to the Chern number used in periodical structures, for all energy gaps with gap value of $\Delta E>0.04 t$. 
	We find that some energy gaps have a nonzero topological invariant $|\nu_{\text{Bott}}|=1$ as indicated by red dots in the third column of Fig.~\ref{fig:resultAB}. 
	We also confirm the existence of edge modes for these nontrivial gaps under the open boundary condition, which are presented in the last column of Fig.~\ref{fig:resultAB}.
	In the fourth column of Fig.~\ref{fig:resultAB}, we show energy levels as a function of $t_d$.
	The green and red dots in these figures show trivial $\nu_{\text{Bott}}=0$ and topological $|\nu_{\text{Bott}}|=1$ gaps, respectively, evaluated by the Bott index.
	In all configurations, a sufficient amount of $t_d$ induces topological states by opening new topological gaps.
	
	We note that STDH is essential in the establishment of topological states.
	For instance, in the fifth column of Fig.~\ref{fig:resultAB}, we calculate the Bott index in the presence of non-staggered tile-diagonal hopping (NSTDH), i.e., $+t_d$ for all tiles with $\pm\pi$ flux.
	Since the value of the Bott index is zero, none of the tiny gaps have a nontrivial topological state.
	
 	\begin{figure}[t]
		\centering
		\includegraphics[width=0.8\linewidth]{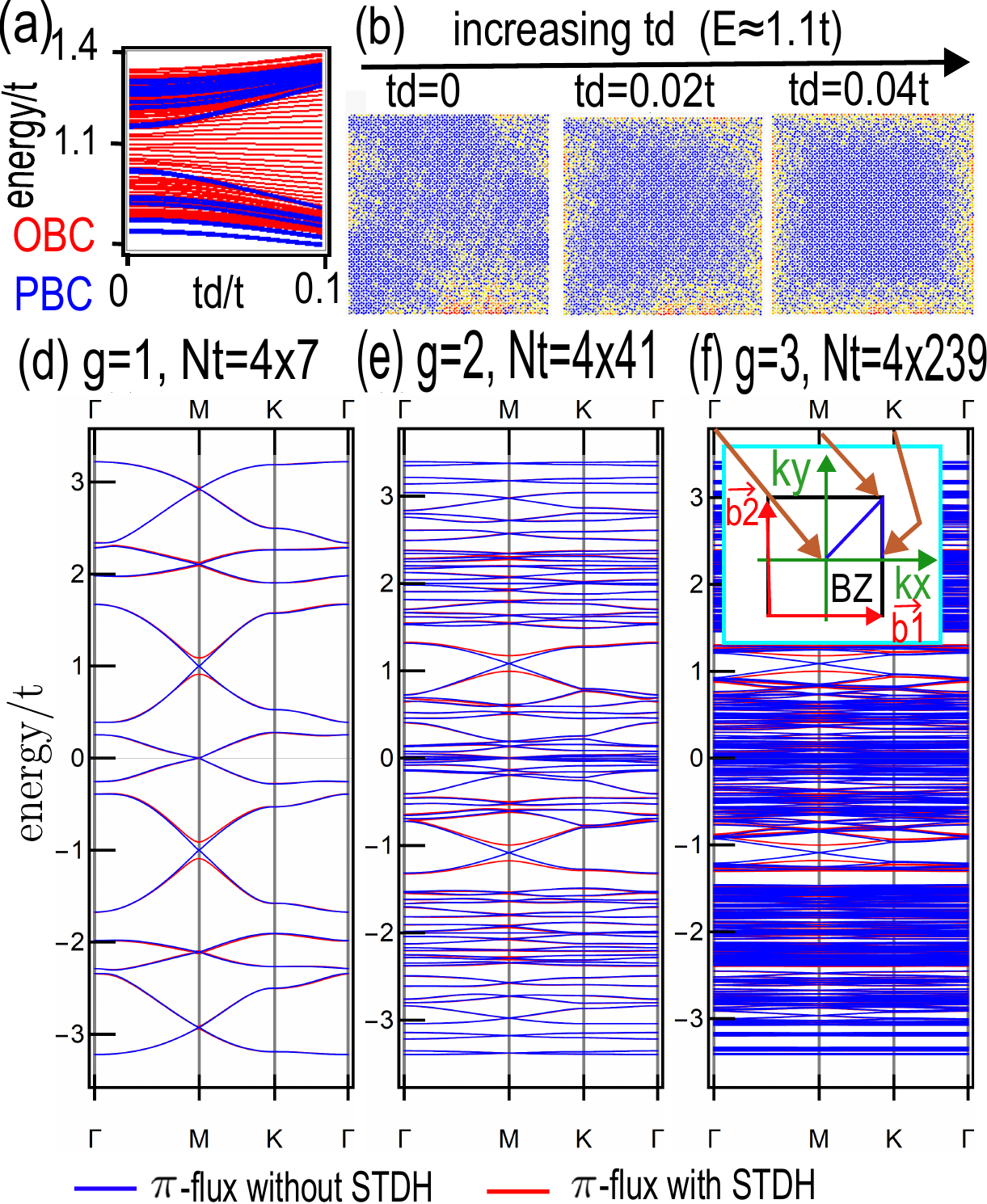}
		\caption[quasicrystal]{
 (a) Energy spectrum with OBC (red lines) and PBC (blue lines) for AB.conf2 as a function of STDH $t_d$.
 (b) The distribution of a wave function with energy $E\approx 1.1 t$ with OBC for different $t_d$. 
 In (a) and (b) we use the same system as we considered in Fig.~\ref{fig:resultAB}(b4).
  (d), (e), (f) Band dispersion of a \piflux AB.conf2 model along high symmetry line for $t_d=0$ (blue) and $t_d=0.1 t$ (red) with different approximant sizes.
			(d) $g=1$, (e) $g=2$, and (f) $g=3$.
			In the inset of (f) we plot Brillouin zone (BZ) and indicate $\vec{k}_{\Gamma}=0$, $\vec{k}_{M}=\tfrac{1}{2}\vec{b}_1+\tfrac{1}{2}\vec{b}_2$, and $\vec{k}_{K}=\tfrac{1}{2}\vec{b}_1$.	}
		\label{fig:DiracCone}
	\end{figure}

		\textit{Dirac cone--}
  	As shown in Fig.~\ref{fig:resultAB}(b4), nonzero STDH in AB.conf2 can induce energy gaps with nontrivial topological Bott index.
		Note that in this case, increasing STDH adiabatically (without gap closing) connects the trivial energy gap (with $v_{\text{Bott}}=0$) located around $E\approx\pm  1.1 t$ with a topological gap (with $v_{\text{Bott}}\ne 0$) [see Fig.~\ref{fig:resultAB}(b4)].  
  In Fig.~\ref{fig:DiracCone}(a), we compare the resulting energy spectrum using periodic boundary conditions (PBC) and open boundary conditions (OBC) and find states inside the gap under PBC irrespective of STDH.
  We plot wave function distribution for different STDH  in Fig.~\ref{fig:DiracCone}(b) and observe that edge modes become visible with increasing $t_d$, leading to clear edge mode in Fig.~\ref{fig:resultAB}(b6). 
  Note that the bulk-boundary correspondence of the topological phases is obtained by considering different twisted boundary conditions. 
  To see this, we examine how energy dispersion (corresponding to different twisted boundary conditions) around $E\approx\pm 1.1 t$ behaves when we take an approximant of AB.conf2 as a unit-cell of translationally invariant system with the basis vectors, $\vec{a}_1$ and $\vec{a}_2$, given in Fig.~\ref{fig:tiling}(b2).
	We apply the Fourier transformation by $\ket{i}=1/N_t\sum_{\vec{k} }\exp(i \vec{r_i}\cdot\vec{k}) \ket{\vec{k}}$ on Eq.~(\ref{eq:APM}), which gives momentum-dependent Hamiltonian $H(\vec{k})$ written by an $N_t\times N_t$ matrix.
	The energy dispersion is obtained by diagonalizing $H(\vec{k})$, where we denote it by $\epsilon_{\lambda=\{1,\dots,N_{t}\}}(\vec{k})$.
	In Fig.~\ref{fig:DiracCone}, we plot $\epsilon_{\lambda}(k)$ with blue (red) color at $t_d=0$ ($t_d=0.1 t$) along high symmetry lines for different approximant size.
	In Fig.~\ref{fig:DiracCone}(a), we find several two-fold degenerate Dirac cones with a liner dispersion at $\vec{k}_{M}=\tfrac{1}{2}\vec{b}_1+\tfrac{1}{2}\vec{b}_2$, where $\vec{b}_1$ and $\vec{b}_2$ are reciprocal vectors.
	The Dirac cones that are located at $E\approx\pm 1.1 t$ become gapped after introducing STDH. 
	This is in accordance with our previous result of the emergence of topological gaps in Fig.~\ref{fig:resultAB}(b4). 
	Increasing system size induces more bands, and therefore, the velocity of these Dirac cones decreases due to the band repulsion from other states.
	Interestingly, the Dirac cones do not repeat with increasing the approximant size, indicating no band folding of the cones.
 
      Note that both the energy gap and system size should be large enough for the calculation of the Bott index \cite{HASTINGS20111699,Toniolo2022}. However, as we found in the previous paragraph, increasing system size decreases the energy gap at $t_d=0$. This means that the calculation of the Bott index at $t_d=0$ is inapplicable to any size of the Ammann-Beenker structure.

		\begin{figure}[t]
		\centering
		\includegraphics[width=0.9\linewidth]{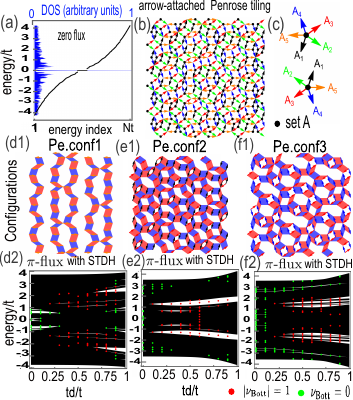}
		\caption[quasicrystal]{ Penrose \piflux model. 
(a) Energy spectrum (black points) and DOS (blue lines) for approximant of Penrose tiling without flux with $g=8$ [$N_t=9349$]. 
(b) Penrose tiling where arrows are attached.
(c) $A_n$ for the Penrose tiling.
(d1), (e1), (f1): Possible Penrose \piflux models.  (d2), (e2), (f2): corresponding energy spectra as a function of STDH $t_d$, where red (green) dots indicate topologically nontrivial (trivial) gaps}
		\label{fig:Penrose}
	\end{figure}
 
	\section{Penrose \piflux model}\label{SEC:PEN}
	The Penrose tiling without \piflux  shows zero energy confined states \cite{PhysRevB.96.214402} (see Fig.~\ref{fig:Penrose}(a)), originating from a local imbalance of bipartite neutrality within a given cluster~\cite{PhysRevB.102.064210}.
	The confined states are separated by a gap and its fraction is given by $p=p_{E=0}=-50\tau_g+81\approx0.098$~\cite{PhysRevB.38.1621,PhysRevB.96.214402}, where $\tau_g=(1+\sqrt{5})/2$ is the golden ratio.
	The imbalance of bipartite neutrality in neighboring clusters alternates between two bipartite sets and is separated by forbidden ladders. 
	As confined states originate from local bipartite imbalance fluctuations, random flux cannot alter the energy and fraction of the original confined states despite significant modification of their wave functions~\cite{PhysRevB.102.064210}. 
	We extend our \piflux approach to the Penrose tiling by defining $A_n$  (see $A_{n=1, \dots,5}$ in Fig.~\ref{fig:Penrose}(c) and their attachment on the edge of the Penrose tiling in Fig.~\ref{fig:Penrose}(b)).
	We confirm that the number of the confined states is unchanged (see Appendix~\ref{sec:confinepenrose}) for any of three possible \piflux configurations (see Figs.~\ref{fig:Penrose}(d1), \ref{fig:Penrose}(e1), and \ref{fig:Penrose}(f1)). 
 \blue{Note that similar to Ammann-Beenker tiling, these configurations are the only possible configuration and others are given by rotation or/and translation of the three configurations.}
	Additionally, we find that introducing sufficient STDH in the Penrose \piflux model gives rise to nontrivial topological states (see Figs.~\ref{fig:Penrose}(d2), \ref{fig:Penrose}(e2), and \ref{fig:Penrose}(f2)).

\section{Summary and Discussion}\label{SEC:SUM}
		In summary, we have extended the two-dimensional \piflux model on a square lattice to quasicrystalline bipartite tilings and made a new proposal for an aperiodic Haldane model by introducing staggered tile-diagonal hopping. 
	We have found that the \piflux model leads to a new massively degenerate confined state in the Ammann-Beenker tiling, while confined states in the Penrose tiling remain the same as the vertex model without flux.
	Additionally, we have discovered massless Dirac cones in a specific \piflux configuration in the Ammann-Beenker approximant.
	Our results on the quasicrystalline \piflux model warrant further exploration through innovative techniques such as topological circuits~\cite{Lv2021_quasicrystalline_quadrupole,Lee2018}, photonics~\cite{Lu2014}.
\blue{For instance, Lv \textit{et al}. \onlinecite{Lv2021_quasicrystalline_quadrupole}  have realized a \piflux distribution in a modified version of the Ammann-Beenker lattice, where each vertex contains four sites. This setup can be easily extended to our Ammann-Beenker and Penrose \piflux models.}
 
	\begin{acknowledgements}
		R.G. was supported by the Institute for Basic Science in Korea (Grant No. IBS-R009-D1),
		Samsung  Science and Technology Foundation under Project Number SSTF-BA2002-06,
		the National Research Foundation of Korea (NRF) grant funded by the Korea government (MSIT) (No.2021R1A2C4002773, and No. NRF-2021R1A5A1032996). M.H. was supported by JST SPRING (Grant No. JPMJSP2151). T.S. was supported by the Japan Society for the Promotion of Science, KAKENHI (Grant No. JP19H05821).
	\end{acknowledgements}

\appendix

 \section{$\pi$-flux model on square lattice}\label{sec:pifluxsquare}
 
  \begin{figure}
 	\centering
 	\includegraphics[width=1\linewidth]{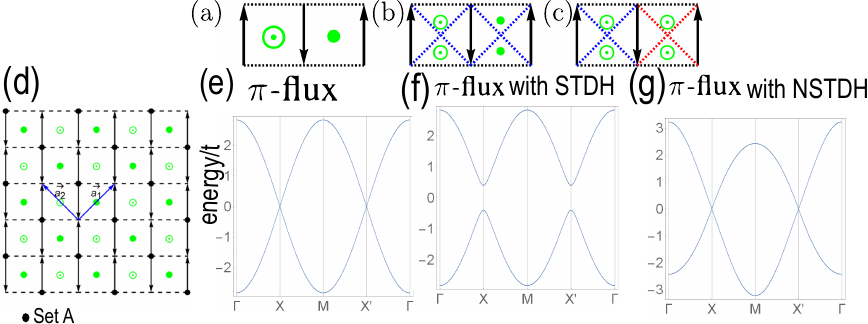}
 	\caption{The $\pi$-flux model in the square lattice.
 		Two tiles of (a) $\pi$-flux, and $\pi$-flux with (b) staggered tile diagonal hopping (STDH), and (c) non-staggered tile diagonal hopping (NSTDH).
 		The two green symbols inside the tiles indicate positive and negative $\pi$ [$\pi/2$] flux in (a) [(b) and (c)].
 		In (c), red and blue diagonal links indicate hopping with opposite signs regarding each other.
 		(d) The $\pi$-flux model in the square lattice, where $\vec{a_1}, \vec{a_2}$ show basis vectors.
 		The energy dispersion for (e) $t_d=0$, (f) STDH with $t_d=0.1t$, and (g) NSTDH with $t_d=0.1t$ along high-symmetry lines.}
 	\label{fig:pifluxsquare}
 \end{figure}

 The  $\pi$-flux model in the square lattice reads 
 \begin{equation}\label{eqpisquare}
 	H=t\sum_{\expval{ij}} e^{iA_{ij}} \ket{i}\bra{j},
 \end{equation}
 where $t$ is the electron hopping. $\ket{i}\bra{j}$ represents hopping operator between site $i$ and $j$, and $\expval{ij}$
 is the nearest neighbor links of the square lattice, or equivalently edges of the square lattice, and $A_{ij}=-A_{ji}$ is the phase of hopping.
 The total flux penetrating each face or tile of the square lattice is given by the total phases that accumulate around edges, i.e.,
 \begin{equation}\label{key}
 	\phi \equiv \phi_{i,j,k,l}=A_{ij}+A_{j,k}+A_{k,l}+A_{l,i}.
 \end{equation}
where $i$, $j$, $k$, and $l$ represent vertices on a given tile and are arranged in an anticlockwise way such as $i\rightarrow j\rightarrow k \rightarrow l \rightarrow i$.
 The square lattice has bipartite properties. This means that we can divide its vertices into two sets, A and B, without any edges connecting the same set.  In such a bipartite Hamiltonian, energy eigenvalues come in pair $\pm E$, leading to a symmetric spectrum.
 We can obtain a $\pi$-flux distribution, by assigning $A_{i,j}=\pi/2$ ($A_{j,i}=-\pi/2$), if $i$ ($j$) belongs to A (B) set and $\vec{r}_i-\vec{r}_j$ is parallel to either horizontal or vertical axis, where $\vec{r_i}$ is the position vector of site $i$.
 In Fig.~\ref{fig:pifluxsquare}(d) we show a $\pi$-flux model in the square lattice, where arrows along the vertical direction represent $A_{i,j}=\pi/2$ belonging to the A set and broken horizontal lines connecting two sites represent $A_{i,j}=0$. Figure~\ref{fig:pifluxsquare}(a) represents two tiles in the $\pi$-flux model.
 
 We can obtain the same $\pi$-flux distribution if we assign $A_{i,j}=\pi/2 (0)$ along horizontal (vertical) directions.
 The energy dispersion of the $\pi$-flux model is given by diagonalizing the Fourier-transformed Hamiltonian
 \begin{equation}\label{dispersionsquarepiflux}
 	H(k)=2t \cos(k_x)\sigma_x+2t \cos(k_y)\sigma_y.
 \end{equation}
 Figure~\ref{fig:pifluxsquare}(e) shows the resulting energy dispersion with two gapless Dirac cones.
 
 The Dirac cones of the square $\pi$-flux model can be gapped out by introducing staggered tile diagonal hopping (STDH)
 \begin{equation}\label{eqpisquare}
 	H'=t_d\sum_{\expval{\expval{ij}}_f} e^{iA_{ij}} \eta_f \ket{i}\bra{j},
 \end{equation}
 where $\expval{\expval{ij}}_f$ represents a pair of diagonal vertices of the faces indicated by $f$ and $\eta_f=1 (-1)$ if the corresponding face has $\pi$ ($-\pi$) flux [see Fig.~\ref{fig:pifluxsquare}(b)]. The Fourier transformation of $H'$ gives
 \begin{equation}\label{key}
 	H'(k)= 	2t_d \sin k_x\sin k_y \sigma_z.
 \end{equation}
 As shown in Fig.~\ref{fig:pifluxsquare}(f), the gap of the Dirac cones opens up by introducing STDH.
 We can confirm the topological state for $t_d\ne0$, by calculating the Chern number which gives nonzero and quantized values.
 Furthermore, due to the bulk-boundary correspondence, the system hosts  chiral edge modes in the open boundary condition.

 Note that introducing non-staggered tile diagonal hopping  (NSTDH) [see Fig.~\ref{fig:pifluxsquare}(c)] does not open up the gap [see Fig.\ref{fig:pifluxsquare}(g)] as the underlying system still has space-time symmetry.

 \section{inflation rule to obtain arrow-attached Ammann-Beenker and Penrose tilings}\label{sec:inflation}
 
\begin{figure*}[!]
 	\centering
 	\includegraphics[width=\linewidth]{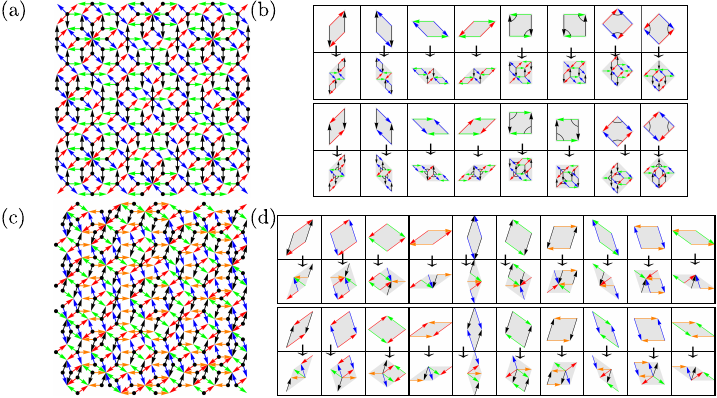}
 	\caption[quasicrystal directional graph]{ (a) Arrow-attached Ammann-Beenker and (c) arrow-attached Penrose tilings. (b) Inflation rule for the Ammann-Beenker tiling and (d) that for the Penrose tiling.}
 	\label{fig:Arrow}
 \end{figure*}

 In Figs.~\ref{fig:Arrow}(a) and \ref{fig:Arrow}(c), we present arrow-attached Ammann-Beenker and Penrose tilings, where four and five distinct arrows exist, respectively.
 Note that, connecting arrows, we obtain a strip of arrows whose adjacent arrows have opposite directions.
 These strips are along a certain direction, which is known as the Ammann bar in quasicrystal structures.
 
 Note that one can obtain an inflation rule to generate these arrow-attached quasicrystals.
 In the inflation rule, each tile is substituted by several combinations of given quasicrystal tiles in a way that quasicrystalline tiling holds after inflation/deflation.
 Therefore, by rescaling new quasicrystalline tiling to have the same tiling as before, one can obtain a quasicrystal with a larger number of tiles.
 To obtain the inflation rule of arrow-attached quasicrystals, we need to know the original and inflated tiles.
 In Figs.~\ref{fig:Arrow}(b) and \ref{fig:Arrow}(d), we show all distinct tiles and their inflated ones for the Ammann-Beenker and Penrose tilings, respectively.

 \section{Bott index}\label{sec:Bott}
 To determine the topological characteristic of the quasicrystalline $\pi$-flux model, we use the Bott index. This index is a real space topological invariant and has been used for systems without time-reversal symmetry.
 This index for a given energy $\epsilon_f$ is defined by
 \begin{equation}
 	B(\epsilon_f)=\frac{1}{2\pi}\Im\left[\text{tr}\log\left(X_{\epsilon_f}^\dagger Y_{\epsilon_f}^\dagger X_{\epsilon_f} Y_{\epsilon_f}\right) \right],
 \end{equation}
 where $\Im$ is the imaginary part, tr is the matrix trace, $\log$ is the matrix log,  $X_{\epsilon_f}=P_{\epsilon_f}XP_{\epsilon_f}+Q_{\epsilon_f}$, and $Y_{\epsilon_f}=P_{\epsilon_f}YP_{\epsilon_f}+Q_{\epsilon_f}$. The projection operators $P_{\epsilon_f}, Q_{\epsilon_f}$ are given by
 \begin{equation}\label{eq:projection}
 	P_{\epsilon_f}{=}\sum_{\epsilon<\epsilon_f} \ket{\epsilon}\bra{\epsilon},\qquad Q_{\epsilon_f}{=}\sum_{\epsilon>\epsilon_f} \ket{\epsilon}\bra{\epsilon},
 \end{equation}
 where $\epsilon$ and $\ket{\epsilon}$ are the eigen-energy and eigen-states of the given Hamiltonian, respectively.
 The modified position matrix is defined by
 \begin{equation}\label{eq:position}
 	\begin{matrix}
 		X{=}\text{diag}\left[e^{2\pi i x_1/L_x}, e^{2\pi i x_2/L_x}, \dots, e^{2\pi i x_N/L_x}\right]\\\\
 		Y{=}\text{diag}\left[e^{2\pi i y_1/L_y}, e^{2\pi i y_2/L_y}, \dots, e^{2\pi i y_N/L_y}\right],
 	\end{matrix}
 \end{equation}
 where \textrm{diag} indicates diagonal matrix, $(x_i,y_i)$ is the coordinates of vertices, and $L_x$ ($L_{y}$) is the length along $x$ ($y$) direction.
 Note that, for the calculation of the Bott index, we have to obtain $\epsilon$ and $\ket{\epsilon}$ with the periodic boundary condition, where we need to use a large approximant.
 Furthermore, the Bott index is not reliable when a given energy gap is small. 
 First, we calculate the Bott index for the gap with $\Delta E>0.04 t$ and then check the existence of edge modes for the open boundary condition.
 To obtain a large approximant with the open boundary condition, we remove inter-approximant links in periodic approximants.

 \section{Finding local confined states}\label{sec:IPR}
 In this section, we explain our strategy to find well-localized confined states.
 Consider general Hamiltonian $\mathcal{H}$.
 We can obtain eigenstate $\ket{\lambda}$ with eigenenergy $\epsilon_{\lambda}$ by numerically diagonalizing $\mathcal{H}$,
 \begin{equation}\label{key}
 	\mathcal{H}\ket{\lambda}=\epsilon_{\lambda}\ket{\lambda}.
 \end{equation}
 The obtained eigenstates are not necessary well localized and generally have spatial overlap among the wavefunctions.
 To obtain well-localized confined states, we first identify all eigenstates that are hugely degenerate with the same  energy value.
 Note that to suppress the effects of boundary conditions, we need to introduce disorder potential for the boundary vertices.
 Searching localized wavefunctions, we need to choose a small portion of the system.
 In this region, we can construct new eigenstates from the degenerate eigenstates, $\ket{\lambda_\alpha}=\sum \phi_{\alpha i}\ket{i}$, where $\alpha=1,\dots, N_c$ indicates the $N_c$-fold degenerate states and $i$ represents indices of site, orbital, spin, etc.
 For this propose, we maximize inverse participation ratio (IPR) for $\ket{\psi}\equiv \sum\psi_{\alpha}\ket{\lambda_{\alpha}}=\sum_i\sum_{\alpha} \psi_{\alpha}\phi_{\alpha i}\ket{i}$, 
 which is defined as
 \begin{equation}\label{key}
 	\text{IPR}=\frac{\sum_i |\sum_{\alpha} \psi_{\alpha}\phi_{\alpha i}|^4}{\sum_i |\sum_{\alpha} \psi_{\alpha}\phi_{\alpha i}|^2}.
 \end{equation}
 It is known that, if IPR is equal to one (zero), $\ket{\psi}$ is localized (extended).
 We numerically optimize $\ket{\psi}$ to achieve the largest IPR.
 Therefore, we can obtain a confined state denoted by $\ket{\psi_1}$.
 Subtracting  $\ket{\psi_1}$ from a set of $\ket{\lambda_\alpha}$, we can find a next confined state $\ket{\psi}_2$ from $N_c-1$ remaining set of independent eigenstates by maximizing IPR.
 We can finally find $N_c$ well-localized confined states in this region by continuing this procedure.
 By redoing the calculation for other parts of the system, we can obtain all distinct confined states.
 Note that the optimization cost increases rapidly with increasing $N_c$.

 \section{The fraction of confined states in Penrose tiling}\label{sec:confinepenrose}
 
  \begin{figure}
 	\centering
 	\includegraphics[width=\linewidth]{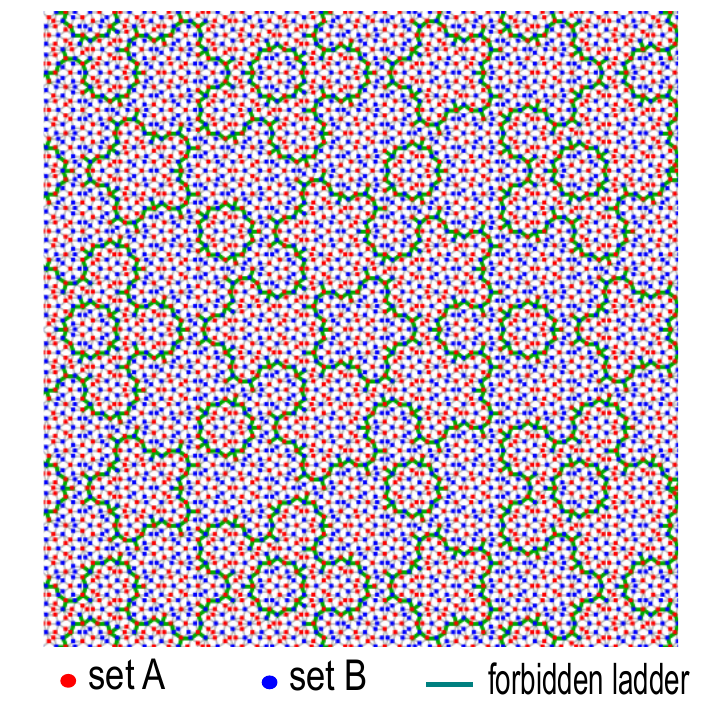}
 	\caption[quasicrystal]{Penrose tiling, where the bipartite index of vertices is given by red and blue dots. The green lines show the forbidden ladder and separate regions with different bipartite imbalances.}
 	\label{fig:penroseConfine}
 \end{figure}

 \subsection{The fraction of confined states}
 Koga and Tsunetsugu~\cite{PhysRevB.96.214402} have calculated the fraction of confined states in Penrose tiling by considering the cluster structures systematically. In Penrose tiling, the confined states have finite amplitudes in  specific regions  named \textit{clusters} in Ref.~\cite{PhysRevB.96.214402}. There are several sizes of clusters [see Fig.~\ref{fig:penroseConfine}], where we label the $i$-th smallest cluster as cluster-$i$. The clusters are surrounded by stripe regions  named \textit{forbidden ladders} in Ref.~\cite{PhysRevB.38.1621}. Koga and Tsunetsugu~\cite{PhysRevB.96.214402} have calculated the generation dependence of the number of confined states in each cluster. The confined states in each cluster have finite amplitudes either in the A sublattice or B sublattice only. Here, following Ref.~\cite{PhysRevB.96.214402}, M and D sites refer to the sites of a sublattice with finite and zero amplitudes, respectively. It has been shown that the number of confined states in each cluster coincides with the difference between the number of M  and D sites in the cluster~\cite{PhysRevB.96.214402}. In other words, the confined states in Penrose tiling are related to the imbalance between A and B sublattices\cite{PhysRevB.102.064210}. This systematic approach reveals the fraction of confined states in the nonzero flux model on Penrose tiling.
 
\begin{figure}
 	\includegraphics[width=0.75\linewidth]{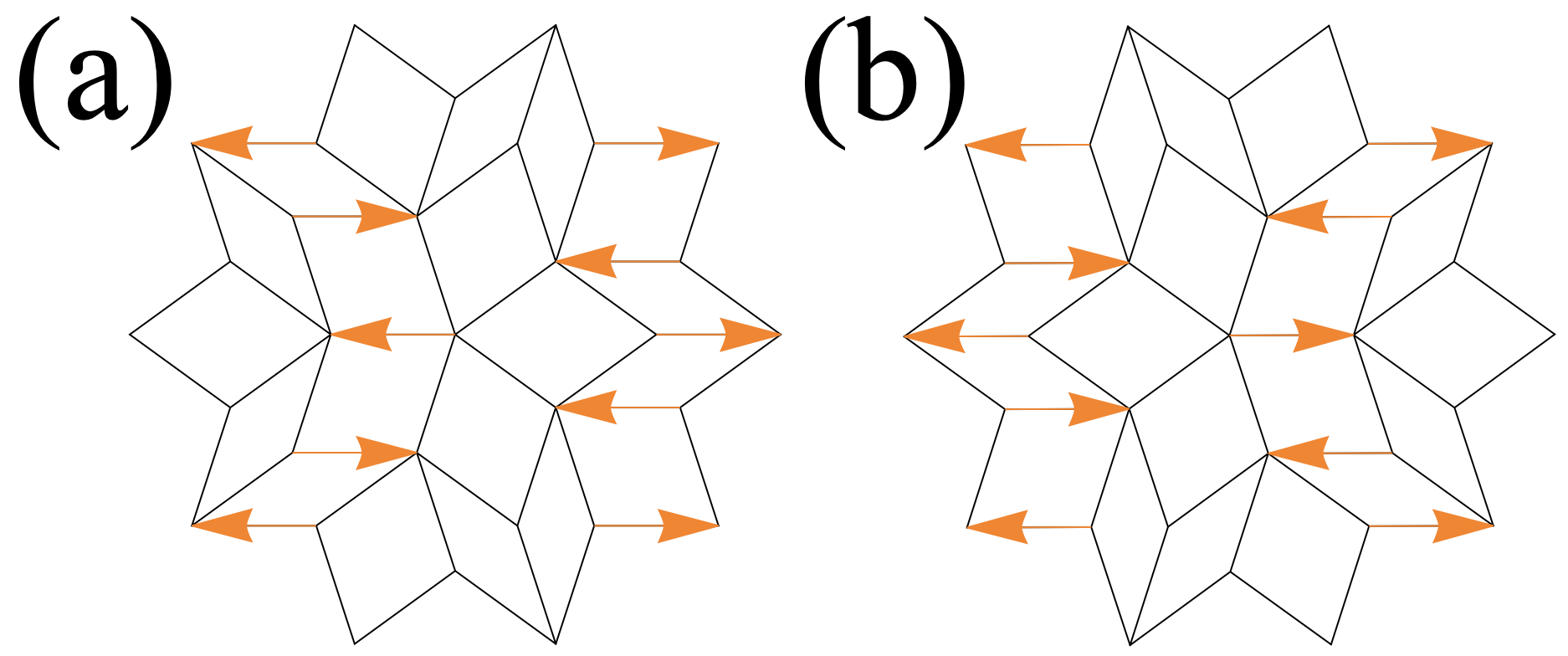}
 	\caption{Cluster-1 of Pe.conf1 for the $\pi$-flux model. The angle of the cluster is different between (a) and (b).
  The arrows denote the hopping with nonzero Peierls phase (see the main text for definition).}
 	\label{fig:twoCluster1}
 \end{figure}

 We apply the method of Ref.~\cite{PhysRevB.96.214402} to the $\pi$-flux model. Here, we should note that clusters in the $\pi$-flux model are the same as those in the zero flux model. For example, let us consider the configuration Pe.conf1 as a $\pi$-flux model [see Fig.~4(d1) in the main text. Two clusters are shown in Figs.~\ref{fig:twoCluster1}(a) and \ref{fig:twoCluster1}(b). These are examples of cluster-1 with different angles. Because the $\pi$-flux model is angle-dependent, the Hamiltonian for the two clusters can be different. However, the five-fold rotational symmetry and mirror symmetry of the clusters assure that the Hamiltonian of the $\pi$-flux model does not depend on their angles. Accordingly,  Fig.~\ref{fig:twoCluster1}(a) coincides with Fig.~\ref{fig:twoCluster1}(b) by a single mirror reflection about $y$-axis or the $\pi$ rotation about $z$-axis.
 
 Calculating the generation dependence of the number of confined states for Pe.conf1, we find that the number of confined states for the $\pi$-flux model is the same as that of the zero flux model for each generation of the cluster. Also, the number of confined states is the same for other configurations such as Pe.conf2 and Pe.conf3 [see Figs.~4(e1) and 4(f1) in the main text]. This suggests that the confined states in the $\pi$-flux model are also related to the sublattice imbalance of the clusters.
 
 \subsection{The wavefunctions of confined states}
\begin{figure}
 	\includegraphics[width=1\linewidth]{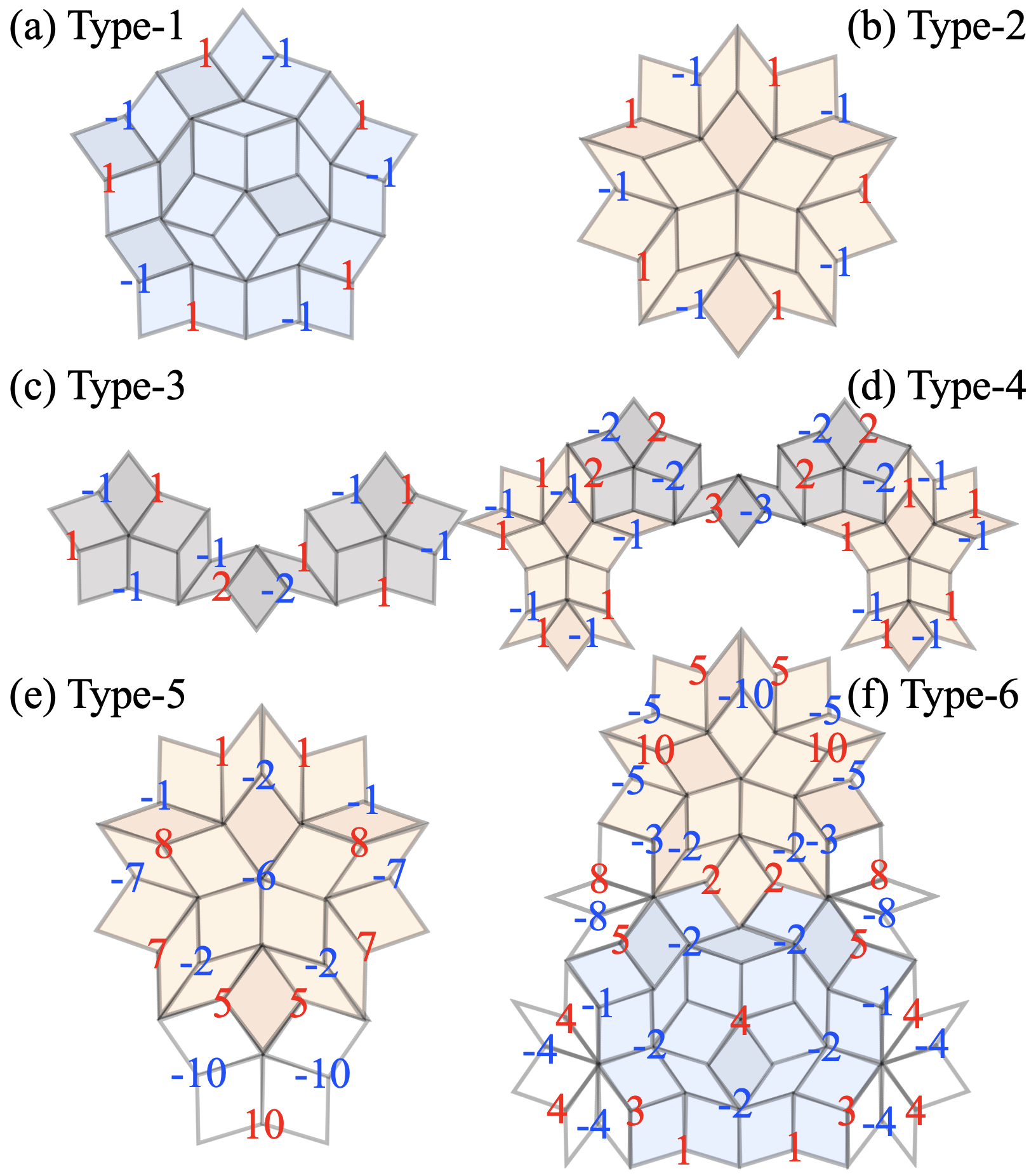}
 	\caption{Six types of confined states in the zero flux model on Penrose tiling. The number on each vertex represents the relative components of the wavefunction.}
 	\label{fig:Penrose_confined_revised}
 \end{figure}

 Next, we consider the wavefunctions of confined states on Penrose tiling. In the case of the zero flux  model, there are only six types of the confined states~\cite{PhysRevLett.56.2740,PhysRevB.38.1621,PhysRevB.96.214402}, illustrated in Figs.~\ref{fig:Penrose_confined_revised}~(a-f). Each number on the vertices represents the relative  components of the wavefunction in each type. The red (blue) color for the numbers implies a positive (negative) sign. Note that the tiles of type-4 and type-5 include those of type-3 and 2, respectively. Also, the tiles of type-6 include those of type-1 and type-2. The number of confined states for each type is summarized in Table.~\ref{table:tightBindingConfType}. 
 
 \begin{table}[]
 	\caption{The number of confined states in the zero flux model for each type in Penrose tiling.}
 	\label{table:tightBindingConfType}
 	\begin{tabular}{l|llllll}
 		\hline\hline
 		type                          & 1 & 2 & 3 & 4 & 5 & 6 \\
 		the number of confined states & 1 & 1 & 1 & 2 & 2 & 3 \\ \hline\hline
 	\end{tabular}
 \end{table}
 
\begin{figure}
 	\includegraphics[width=0.65\linewidth]{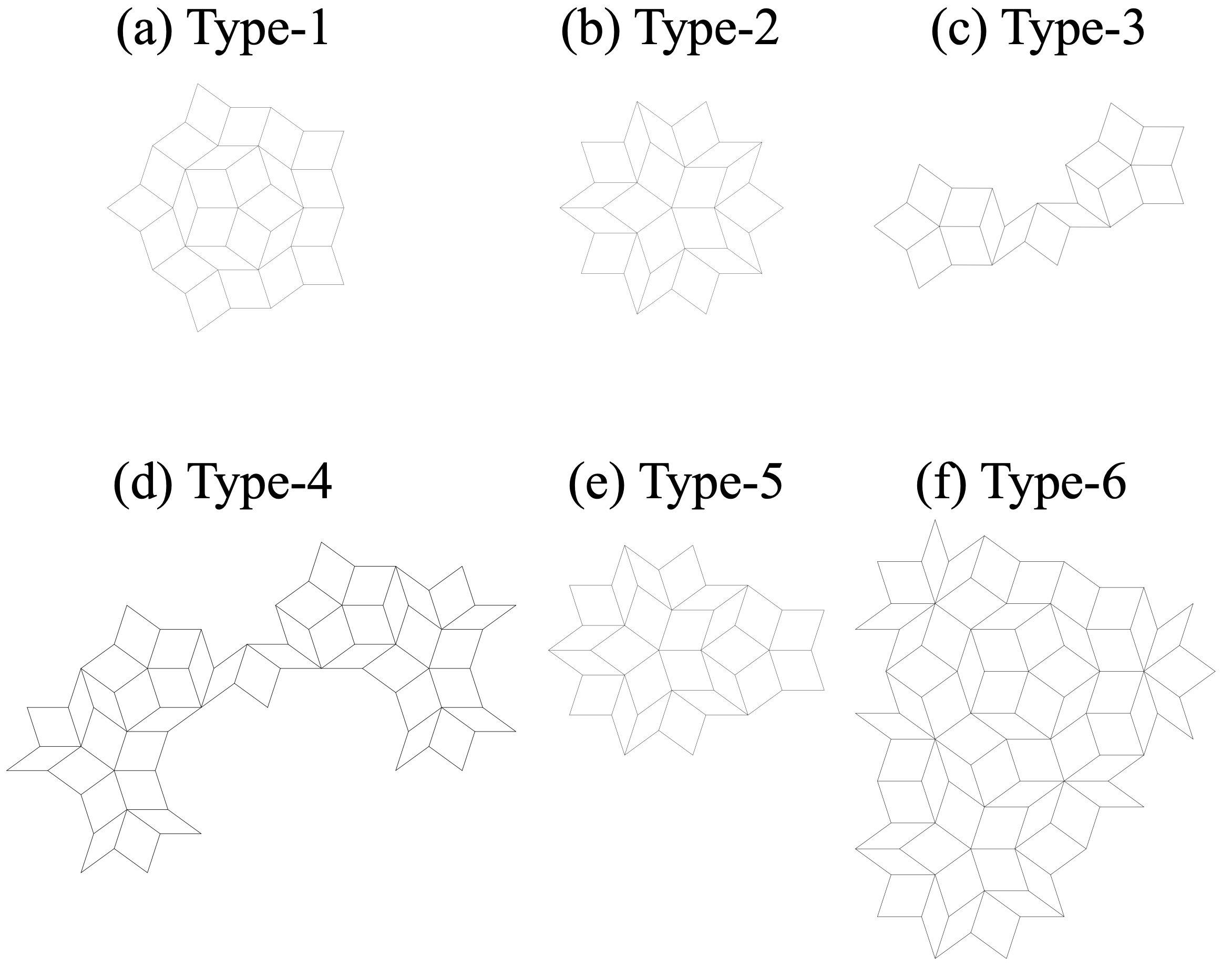}
 	\caption{Six types of the tiles with $r=0$ on Penrose tiling.}
 	\label{fig:PenroseRot0Figs}
 \end{figure}

\begin{figure}
 	\includegraphics[width=0.75\linewidth]{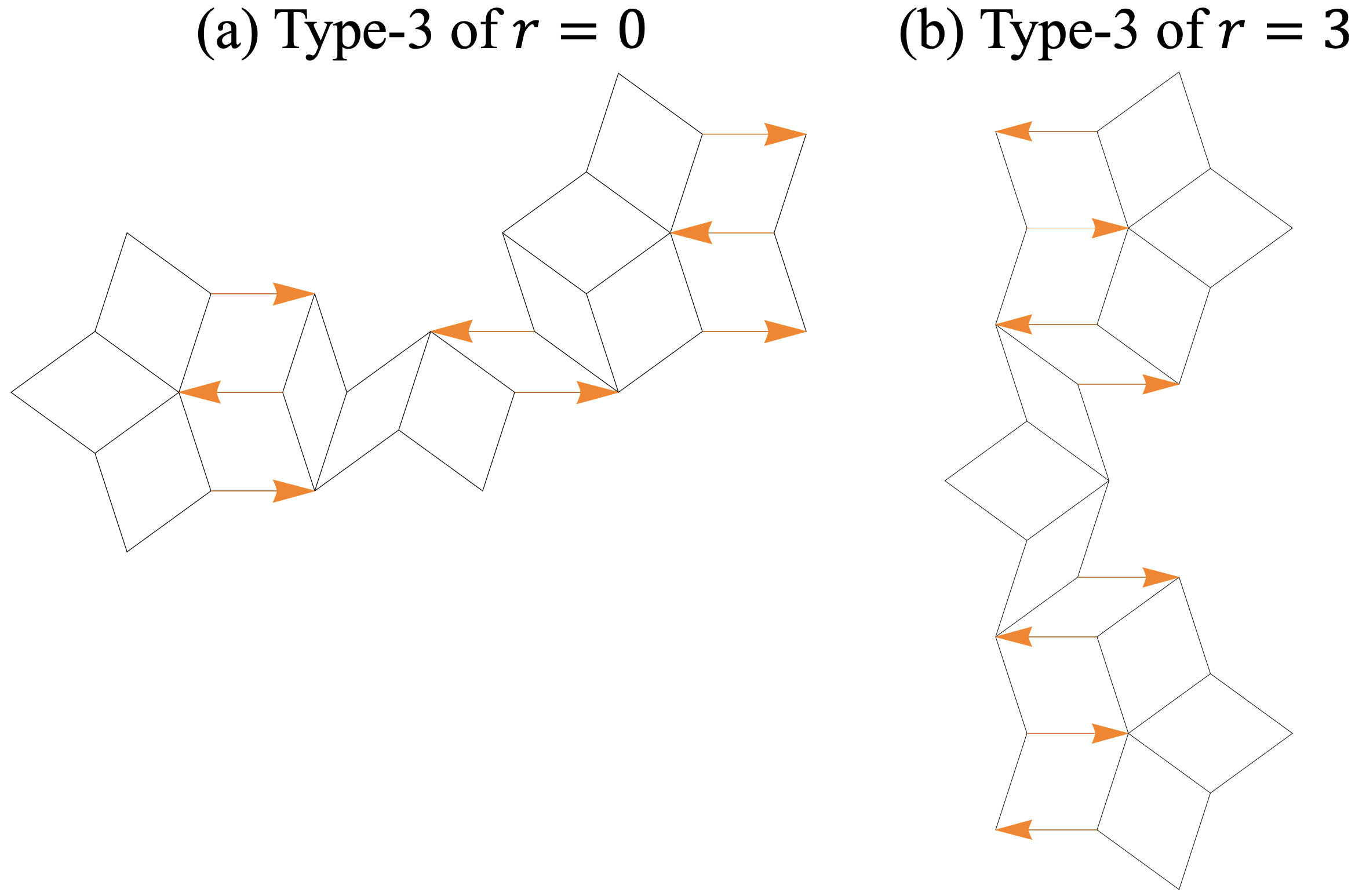}
 	\caption{The type-3 tiles of Pe.conf1in the $\pi$-flux model. (a) $r=0$ and (b) $r=3$. The arrows denote the hopping with nonzero Peierls phase (see the main text for definition).}
 	\label{fig:conf1Type3Rot01}
 \end{figure}

 The confined states in the $\pi$-flux model cannot be classified by the six types shown in Figs.~\ref{fig:Penrose_confined_revised}~(a-f), although the fraction of the confined states in the $\pi$-flux model is the same as that of in the zero flux model.  Due to the lack of five-fold rotational symmetry and mirror symmetry in the $\pi$-flux model, the angle of the tiles is crucial for the type-3,4,5, and 6 in Fig.~\ref{fig:Penrose_confined_revised}. We indicate the angle of the tiles by using a rotational index $r\in{0,1,2,3,4}$. We define the tiles with $r=0$ as illustrated in Figs.~\ref{fig:PenroseRot0Figs}. The tiles with $r$ are obtained by rotating anticlockwise the $r=0$ tiles by $-2\pi r/10$. For example, the type-3 tiles of Pe.conf1 in the $\pi$-flux model are illustrated in Figs.~\ref{fig:conf1Type3Rot01}(a) and \ref{fig:conf1Type3Rot01}(b) for $r=0$ and 3, respectively. 
 
 The number of confined states for a given type and $r$ in several configurations are listed in Table.~\ref{table:numOfConfinedStatesPiAll}. It is clearly seen that the number of confined states for the tiles of type-2,4,5 and 6 in Pe.conf1-3 is the same as those of the zero flux case. We note that the  absence of confine states for type-1 and type-3 does not affect the fraction of the confined states on the entire Penrose tiling, because the tiles of type-1 (type-3) are included in those of type-6 (5 and 6).
 
 \begin{table}[]
 	\caption{The number of confined states for given type and $r$ in the three configurations in the $\pi$-flux model and in zero flux case.}
 	\label{table:numOfConfinedStatesPiAll}
 	\begin{tabular}{lllllll}
 		\multicolumn{7}{l}{(a) Pe.conf1}                                      \\ \hline
 		&                        & \multicolumn{5}{c}{$r$} \\ \cline{3-7}
 		&                        & 0   & 1  & 2  & 3  & 4  \\ \cline{3-7}
 		\multicolumn{1}{l|}{}     & \multicolumn{1}{l|}{1} & 0   & 0  & 0  & 0  & 0  \\
 		\multicolumn{1}{l|}{}     & \multicolumn{1}{l|}{2} & 1   & 1  & 1  & 1  & 1  \\
 		\multicolumn{1}{l|}{type} & \multicolumn{1}{l|}{3} & 0   & 0  & 0  & 1  & 0  \\
 		\multicolumn{1}{l|}{}     & \multicolumn{1}{l|}{4} & 2   & 2  & 2  & 2  & 2  \\
 		\multicolumn{1}{l|}{}     & \multicolumn{1}{l|}{5} & 2   & 2  & 2  & 2  & 2  \\
 		\multicolumn{1}{l|}{}     & \multicolumn{1}{l|}{6} & 3   & 3  & 3  & 3  & 3  \\ \hline
 			\end{tabular}\qquad
 			\begin{tabular}{lllllll}
 		\multicolumn{7}{l}{(b) Pe.conf2}                                      \\ \hline
 		&                        & \multicolumn{5}{c}{$r$} \\ \cline{3-7}
 		&                        & 0   & 1  & 2  & 3  & 4  \\ \cline{3-7}
 		\multicolumn{1}{l|}{}     & \multicolumn{1}{l|}{1} & 0   & 0  & 0  & 0  & 0  \\
 		\multicolumn{1}{l|}{}     & \multicolumn{1}{l|}{2} & 1   & 1  & 1  & 1  & 1  \\
 		\multicolumn{1}{l|}{type} & \multicolumn{1}{l|}{3} & 0   & 0  & 0  & 0  & 1  \\
 		\multicolumn{1}{l|}{}     & \multicolumn{1}{l|}{4} & 2   & 2  & 2  & 2  & 2  \\
 		\multicolumn{1}{l|}{}     & \multicolumn{1}{l|}{5} & 2   & 2  & 2  & 2  & 2  \\
 		\multicolumn{1}{l|}{}     & \multicolumn{1}{l|}{6} & 3   & 3  & 3  & 3  & 3  \\ \hline
 			\end{tabular}\qquad
 		\begin{tabular}{lllllll}
 		\multicolumn{7}{l}{(c) Pe.conf3}                                      \\ \hline
 		&                        & \multicolumn{5}{c}{$r$} \\ \cline{3-7}
 		&                        & 0   & 1  & 2  & 3  & 4  \\ \cline{3-7}
 		\multicolumn{1}{l|}{}     & \multicolumn{1}{l|}{1} & 0   & 0  & 0  & 0  & 0  \\
 		\multicolumn{1}{l|}{}     & \multicolumn{1}{l|}{2} & 1   & 1  & 1  & 1  & 1  \\
 		\multicolumn{1}{l|}{type} & \multicolumn{1}{l|}{3} & 0   & 1  & 0  & 0  & 0  \\
 		\multicolumn{1}{l|}{}     & \multicolumn{1}{l|}{4} & 2   & 2  & 2  & 2  & 2  \\
 		\multicolumn{1}{l|}{}     & \multicolumn{1}{l|}{5} & 2   & 2  & 2  & 2  & 2  \\
 		\multicolumn{1}{l|}{}     & \multicolumn{1}{l|}{6} & 3   & 3  & 3  & 3  & 3  \\ \hline
 	\end{tabular}\qquad
  \begin{tabular}{lllllll}
 		\multicolumn{7}{l}{(d) no flux}                                      \\ \hline
 		&                        & \multicolumn{5}{c}{$r$} \\ \cline{3-7}
 		&                        & 0   & 1  & 2  & 3  & 4  \\ \cline{3-7}
 		\multicolumn{1}{l|}{}     & \multicolumn{1}{l|}{1} & 1   & 1  & 1  & 1  & 1  \\
 		\multicolumn{1}{l|}{}     & \multicolumn{1}{l|}{2} & 1   & 1  & 1  & 1  & 1  \\
 		\multicolumn{1}{l|}{type} & \multicolumn{1}{l|}{3} & 1   & 1  & 1  & 1  & 1  \\
 		\multicolumn{1}{l|}{}     & \multicolumn{1}{l|}{4} & 2   & 2  & 2  & 2  & 2  \\
 		\multicolumn{1}{l|}{}     & \multicolumn{1}{l|}{5} & 2   & 2  & 2  & 2  & 2  \\
 		\multicolumn{1}{l|}{}     & \multicolumn{1}{l|}{6} & 3   & 3  & 3  & 3  & 3  \\ \hline
 	\end{tabular}
 \end{table}
 
 Except for type-3, the number of the confined states in the $\pi$-flux model does not depend on configurations and rotation indexes. Because type-3 has mirror symmetry about the diagonal line of the central tile, the confined states exist only when this mirror symmetry is unchanged even in the presence of $\pi$-flux. Such a symmetry condition is satisfied only for one value of $r$ in each configuration.
 
\begin{figure*}
 	\includegraphics[width=\linewidth]{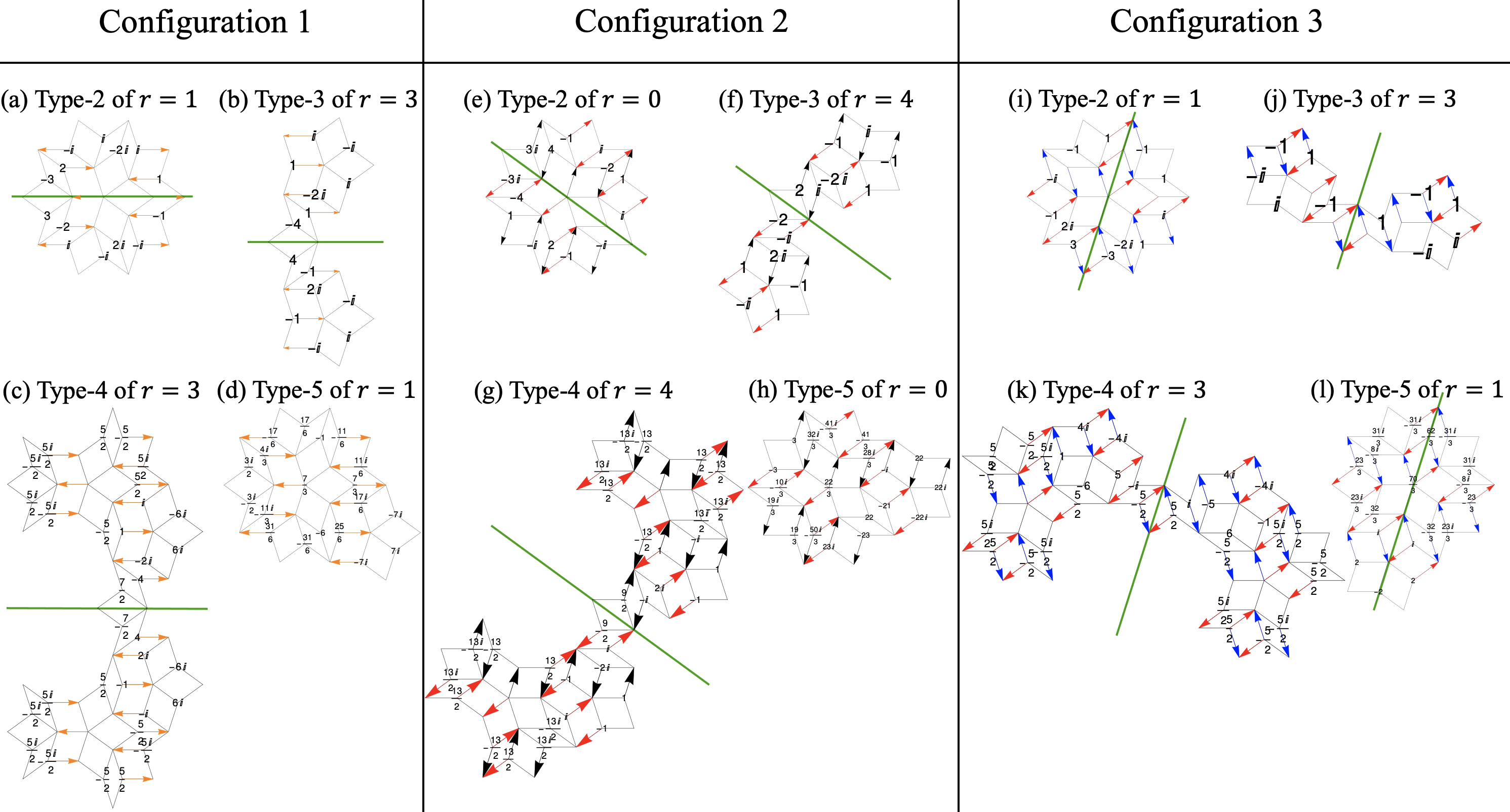}
 	\caption{Some examples of the confined states in Pe.conf1 (a-d), Pe.conf2 (e-h), and Pe.conf3 (i-l) of the $\pi$-flux model on Penrose tiling. The tiles are (a) type-2 with $r=1$, (b) type-3 with $r=3$, (c) type-4 with $r=3$, (d) type-5 with $r=1$, (e) type-2 with $r=0$, (f) type-3 with $r=4$, (g) type-4 with $r=4$, (h) type-5 with $r=0$, (i) type-2 with $r=1$, (j) type-3 with $r=3$, (k) type-4 with $r=3$, and (l) type-5 of $r=1$. Each number written on the lattice sites is the relative  components of the wavefunctions for the confined states with zero energy. The arrows denote the hopping with nonzero Peierls phase (see the main text for definition).}
 	\label{fig:configurationAll_CSs_Examples}
 \end{figure*}

 In Fig.~\ref{fig:configurationAll_CSs_Examples}, we show some examples of the confined states for the given type of tiles and $r$ in each configuration of the $\pi$-flux model in Penrose tiling. 
Note that the confined states for the type-1 do not exist. The confined states for the type-6 are not shown in Fig.~\ref{fig:configurationAll_CSs_Examples}, since they are too complicated to illustrate. There exist two confined states for type-4 and type-5. The tiles in type-4 and 5 include those in type-3 and 2, respectively. Accordingly, one of the confined states in type-4 and type-5 is those in type-3 [Figs.~\ref{fig:configurationAll_CSs_Examples}~(b,f,j)] and type-2 [Figs.~\ref{fig:configurationAll_CSs_Examples}~(a,e,i)], respectively. We select a linear combination of the confined states so that the other confined states in type-4 and type-5 [Figs.~\ref{fig:configurationAll_CSs_Examples}~(c,g,k) and (d,h,l)] are orthogonal to the confined states in type-3 and type-2. The green lines  in Figs.~\ref{fig:configurationAll_CSs_Examples}~(a-c), (d-g), and (i-l) represent the axis of mirror symmetry in each configuration. The confined states shown in Figs.~\ref{fig:configurationAll_CSs_Examples}~(a-c), (d-g), and (i-k) have the odd mirror symmetry about the green lines, while the confined state shown in Fig.~\ref{fig:configurationAll_CSs_Examples}~(l) has the even mirror symmetry about the green line. The $\pi$-flux models illustrated in Figs.~\ref{fig:configurationAll_CSs_Examples}~(d,h) have no mirror symmetry. 
 
 \section{The fraction of confined states in Ammann-Beenker tiling}\label{sec:confineAmmann}
 
 
 Koga~\cite{PhysRevB.102.115125} has calculated the fraction of confined states in Ammann-Beenker tiling by focusing on the domain structures with locally eight-fold rotational symmetry. Koga has considered the zero flux model in Ammann-Beenker tiling, whose fraction of confined states has been obtained analytically as $1/2\tau_{s}^{2}$. In this subsection, we estimate the fraction of confined states in the $\pi$-flux model by following the method utilized in Ref.~ \cite{PhysRevB.102.115125}. 
 
\begin{figure}
 	\includegraphics[width=1\linewidth]{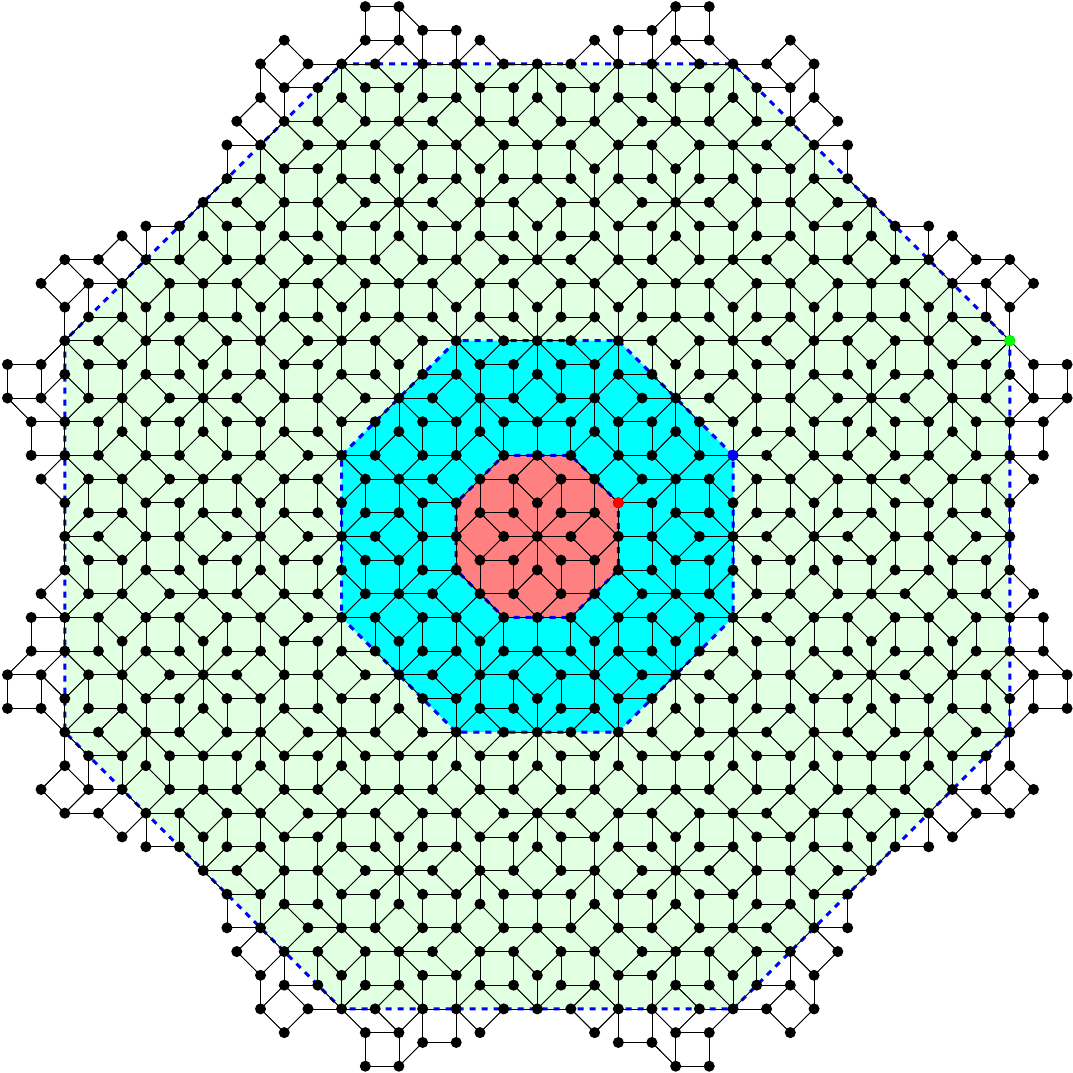}
 	\caption{Ammann-Beenker tiling. The red, blue, and green shaded areas are $D^{1}$, $D^{2}$, and $D^{3}$, respectively. The sites on the boundary that are indicated by blue dashed lines are not included in the domains. For example, the red, blue, and green dots at the boundary of $D^{1}$, $D^{2}$, and $D^{3}$ are excluded.} \label{fig:MH_Koga_Domain}
 \end{figure}

 First, we briefly explain the method used in Ref.~ \cite{PhysRevB.102.115125}. To estimate the fraction of confined states, the number of confined states is systematically counted focusing on the domain structures $D^{1}, D^{2}, D^{3}, \ldots$, illustrated in Fig.~\ref{fig:MH_Koga_Domain}. Here, $D^{i+1}$ is generated by applying inflation/deflation operation to $D^{i}$ for any generation $i\in\mathbb{N}$. For each domain, the boundary sites are excluded. Consequently, the number of sites in $D^{1}, D^{2}$, and $D^{3}$ are 17, 121, 753, respectively. There are sixteen $D^{1}$ structures in $D^{3}$. Note that, when $D^{i}$ and $D^{j}$ ($i>j$) structures share their centers, only $D^{i}$ structure is counted as domain structures and $D^{j}$ is not counted for the sake of preventing a double count. Following this rule, the $D^{1}$ structure around the center of $D^{3}$ is not regarded as $D^{1}$.
 
 The net number of confined states in $D^{i}$ is defined as
 \begin{equation}
 	\label{eq:net_num_of_CSs}
 	N^{{\rm{net}}}_{i} = N^{{\rm{tot}}}_{i} - \sum_{j=1}^{i-1}N_{i,j}N^{{\rm net}}_{j},
 \end{equation}
 where $N^{{\rm{tot}}}_{i}$ is the total number of confined states in $D^{i}$ and $N_{i,j}$ is the number of $D^{j}$ in $D^{i}$. Note that $N^{{\rm{net}}}_{i}$ is the contribution of $D^{i}$ to $N^{{\rm{tot}}}_{j}$ ($i<j$). As a consequence, we can break down $N^{{\rm{tot}}}_{j}$ into the contributions from each domain $i (\leq j)$, which enables us to count the number of confined states systematically.
 
 Since $N_{i,j}$ satisfies $N_{i,j}=N_{i-1,j-1}$, Eq.~\ref{eq:net_num_of_CSs} can be rewritten as
 \begin{equation}
 	\label{eq:net_num_of_CSs2}
 	N^{{\rm{net}}}_{i} = N^{{\rm{tot}}}_{i} - \sum_{j=1}^{i-1}N_{i-j+1,1}N^{{\rm net}}_{j}.
 \end{equation}
 We can numerically obtain $N^{{\rm{tot}}}_{i}$ by diagonalizing the zero flux Hamiltonian. Also, Koga  \cite{PhysRevB.102.115125} has obtained $N_{i,1}$ analytically for $i\in\mathbb{N}$. Consequently, we can calculate $N^{{\rm{net}}}_{i}$ from \ref{eq:net_num_of_CSs2}. Table~\ref{table:conf0NiNet} lists $N_{i1}$, $N^{{\rm tot}}_{i}$, and $N^{{\rm net}}_{i}$ for each generation $i$ ($1\leq i \leq 7$). 
 For instance, in the original Ammann-Beenker tiling   $N^{{\rm{net}}}_{i}=i(i+1)$ (see Table~\ref{table:conf0NiNet}). The fraction of confined states is obtained by
 \begin{equation}
 	\label{eq:p_Koga}
 	p=\sum_{i}p_{i}N^{{\rm{net}}}_{i},
 \end{equation}
 where $p_{i}$ is the fraction of $D^{i}$. Since Koga~\cite{PhysRevB.102.115125} has obtained $p_{i}$ analytically as $p_{i}=2 \tau_{s}^{-(2i+3)}$, we can obtain $p$ analytically when the analytical assumption of $N^{{\rm{net}}}_{i}$ is provided. In the case of the zero flux model, $p$ results in $1/2\tau_{s}^{2}$. We apply Koga's method to our $\pi$-flux models.  
 In the following, we consider each configuration separately. 
 
  Note that, since in the \piflux models, fluxes penetrating tiles are either $\pm \pi$ or zero,  the system has time-reversal symmetry $\mathcal{T} H^* -H  \mathcal{T} =0$. On the other hand, due to the bipartite properties of the structure, the system has chiral symmetry $\mathcal{C} H +H  \mathcal{C} =0$. 
 The product of time-reversal and chiral operators gives a particle-hole operator  $\mathcal{P}=\mathcal{C}\mathcal{T}$, where $\mathcal{P} H^* +H  \mathcal{P} =0$.
 By applying particle-hole (chiral) operation on a wavefunction with energy $E$, one can see  $\Psi\rightarrow \mathcal{P} \Psi^*   $ ($\Psi\rightarrow \mathcal{C} \Psi$) and $E\rightarrow-E$.
 Therefore, in the following, we consider the confined states whose energy is greater than or equal to 0.

 \begin{table}[]
 	\caption{$N_{i1}$, $N^{{\rm tot}}_{i}$, and $N^{{\rm net}}_{i}$ for $1\leq i \leq 7$ in the zero flux  model of Ammann-Beenker tiling.}
 	\label{table:conf0NiNet}
 	\begin{tabular}{llll}
 		\hline\hline
 		$i$ & $N_{i1}$ & $N^{{\rm tot}}_{i}$ & $N^{{\rm net}}_{i}$ \\ \hline
 		1   & 1        & 2                   & 2                   \\
 		2   & 0        & 6                   & 6                   \\
 		3   & 16       & 44                  & 12                  \\
 		4   & 104      & 324                 & 20                  \\
 		5   & 632      & 2110                & 30                  \\
 		6   & 3768     & 12938               & 42                  \\
 		7   & 22152    & 77112               & 56                  \\ \hline\hline
 	\end{tabular}
 \end{table}
 
 \subsection{AB.conf1 with $E=\sqrt{2}$}
 For the AB.conf1 in the $\pi$-flux model [see Fig.~2(a1) in the main text] with $E=\sqrt{2}$, $N_{i1}$, $N^{{\rm tot}}_{i}$, and $N^{{\rm net}}_{i}$ for $1 \leq i \leq 5$ are listed in Table~\ref{table:conf1E1NiNetAPP}. As is clearly shown in Table~\ref{table:conf1E1NiNetAPP},   $N^{{\rm net}}_{i}=1$  for any generation $i$. As a consequence, the fraction of confined states calculated from Eq.~(\ref{eq:p_Koga}) is $p=\tau_{s}^{-4}$ which coincides with $p^{{\rm F}}=\tau_{s}^{-4}$, where $p^{{\rm \alpha}}$ is the fraction of $\alpha$ vertex with $\alpha=$ A, B, C, D, E, and F whose coordination number is 3, 4, 5, 6, 7, and 8, respectively~\cite{PhysRevB.42.8091}.
 
 By definition, $D^{i}$ always includes $D^{i-1}$ structure, where $D^{i}$ and $D^{i-1}$ share their centers. Note that we ignore the $D^{i-1}$ structure when we count the number of $D^{i-1}$ in $D^{i}$. Accordingly, the set of confined states in $D^{i}$ always includes that of in $D^{i-1}$. As a result, the number of confined states in $D^{i}$ is greater than or equal to that of in $D^{i-1}$, namely, $N^{{\rm net}}_{i}-N^{{\rm net}}_{i-1} \geq 0$. The number of confined states that do exist in $D^{i}$ and do not exist in $D^{i-1}$ is $N^{{\rm net}}_{i}-N^{{\rm net}}_{i-1}$.
 In Table~\ref{table:conf1E1NiNetAPP}, $N^{{\rm net}}_{i}-N^{{\rm net}}_{i-1} = 0$ for any $i\in\mathbb{N}$. This implies that all of the confined states in $D^{i}$ for $2\leq i$ are the confined states in $D^{1}$.
 After all, the fact that $N^{{\rm net}}_{i}=1$ for any $i$ implies the presence of only one type of confined state localized inside $D^{1}$, which justifies $p=p^{{\rm F}}$.
 
 \begin{table}[]
 	\caption{$N_{i1}$, $N^{{\rm tot}}_{i}$, and $N^{{\rm net}}_{i}$ for $1\leq i \leq 5$ in AB.conf1 of the $\pi$-flux model on Ammann-Beenker tiling for $E=\sqrt{2}$.}
 	\label{table:conf1E1NiNetAPP}
 	\begin{tabular}{lll}
 		\hline\hline
 		$i$ & $N^{{\rm tot}}_{i}$ & $N^{{\rm net}}_{i}$ \\ \hline
 		1   & 1                   & 1                   \\
 		2   & 1                   & 1                   \\
 		3   & 17                  & 1                   \\
 		4   & 121                 & 1                   \\
 		5   & 753                 & 1                   \\ \hline\hline
 	\end{tabular}
 \end{table}
 
 We consider the confined state in $D^{1}$. The choice of confined states is not unique because a linear combination of confined states is also the eigenstates with the same eigenenergy. Koga~\cite{PhysRevB.102.115125} selected confined states so that they are described by the one-dimensional representation of the dihedral group $D_{8}$. In the case of the $\pi$-flux model of AB.conf1, all of the confined states are described by the irreducible representation of the dihedral group $D_{2}$. A character table for the dihedral group $D_{2}$ is shown in Table~\ref{table:D2char}. We use Table~\ref{table:D2char} to classify the confined states in AB.conf1 of the $\pi$-flux model.
 
 \begin{table}[]
 	\caption{A character table for the dihedral group $D_{2}$, where $E$ is the identity operator and $C_{2}(\alpha)$ is the rotation operator of $\pi$ around $\alpha(=x,y,z)$ axis. Here, the unit vectors for $x$ and $y$ axes are $\mbox{\boldmath $e$}_{x}=(1,0)$ and $\mbox{\boldmath $e$}_{y}=(0,1)$, respectively. The unit vector for $z$ axis, $\mbox{\boldmath $e$}_{z}$, is defined so that $\mbox{\boldmath $e$}_{x}, \mbox{\boldmath $e$}_{y}$, and $\mbox{\boldmath $e$}_{z}$ form the right-handed coordinate system. The arrows denote the hopping with nonzero Peierls phase (see the main text for definition).}
 	\label{table:D2char}
 	
 	\begin{tabular}{lllll}
 		\hline\hline
 		& $E$ & $C_{2}(z)$ & $C_{2}(y)$ & $C_{2}(x)$ \\ \hline
 		$\rm{A}$     & +1  & +1         & +1         & +1         \\
 		$\rm{B_{1}}$ & +1  & +1         & -1         & -1         \\
 		$\rm{B_{2}}$ & +1  & -1         & +1         & -1         \\
 		$\rm{B_{3}}$ & +1  & -1         & +1         & -1         \\ \hline\hline
 	\end{tabular}
 \end{table}
 
\begin{figure}
 	\includegraphics[width=0.35\linewidth]{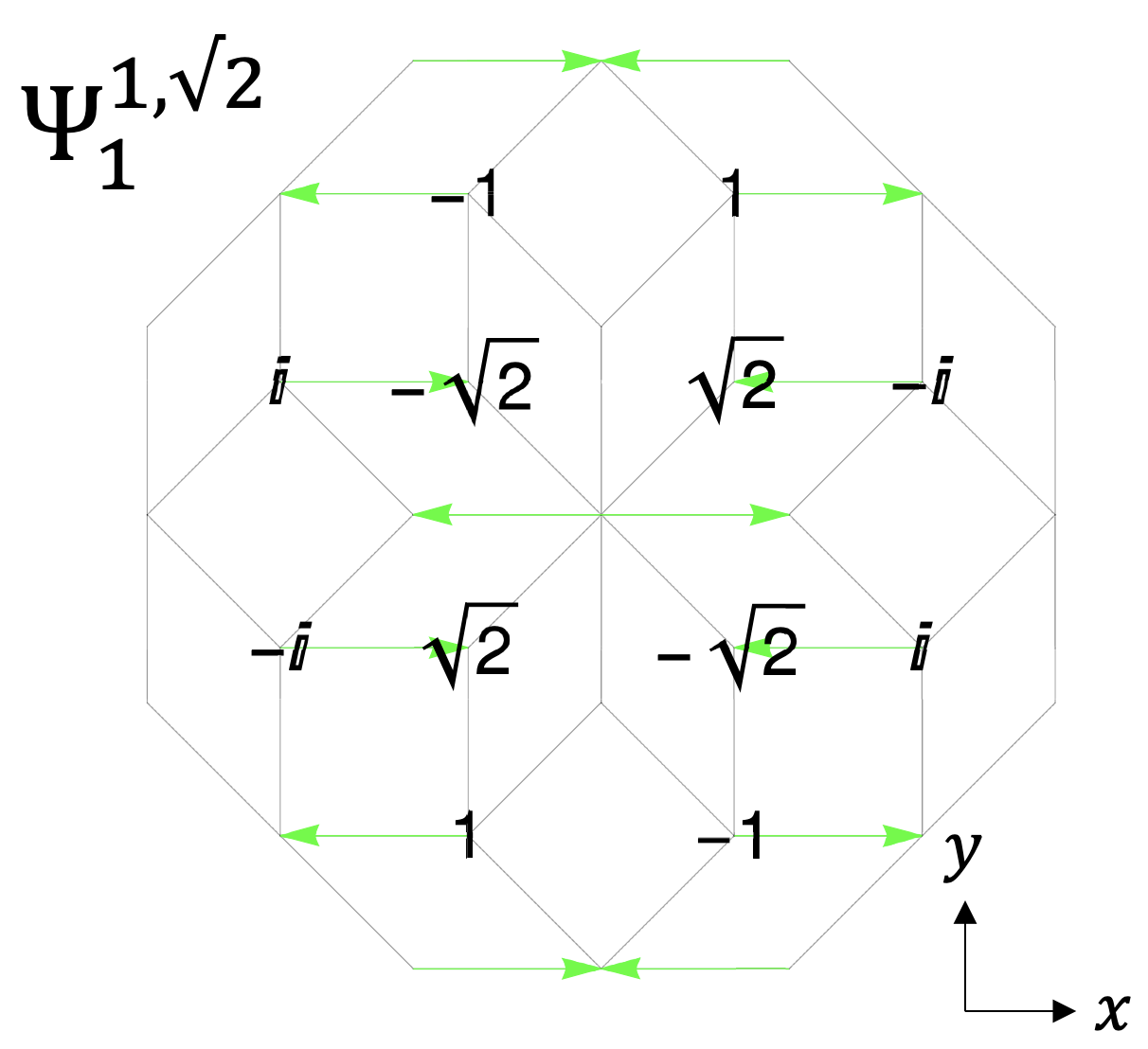}
 	\caption{The confined state in $D^{1}$ for AB.conf1 of the $\pi$-flux model with  $E=\sqrt{2}$ on Ammann-Beenker tiling. The number represents the wavefunction's components of the confined state on each vertex.  The arrows denote the hopping with nonzero Peierls phase (see the main text for definition).}
 	\label{fig:conf1E1CS1}
 \end{figure}

 The confined state in $D^{1}$ is schematically shown in Fig.~\ref{fig:conf1E1CS1}. We denote $\Psi^{c,E}_{k}$ as the confined state of configuration $c$ in the $\pi$-flux model with energy $E$, where $k\in\mathbb{N}$ is the index of confined states. The $\Psi^{1,\sqrt{2}}_{1}$ is a confined state because it is a locally distributed eigenstate regardless of the outside structure of $D^{1}$.
 According to the Table~\ref{table:D2char}, $\Psi^{1,\sqrt{2}}_{1}$ is described by the irreducible representation ${\rm{B_{1}}}$.
 We  note that $\Psi^{1,\sqrt{2}}_{1}$ has amplitudes in both sublattices A and B. 
 As Koga mentioned~\cite{PhysRevB.102.115125}, the confined states in the zero flux model in Ammann-Beenker tiling have amplitudes only in one of the sublattices A and B. This noteworthy feature of Ammann-Beenker tiling is confirmed in AB.conf1 of the $\pi$-flux model.
 
\begin{figure}
 	\includegraphics[width=1\linewidth]{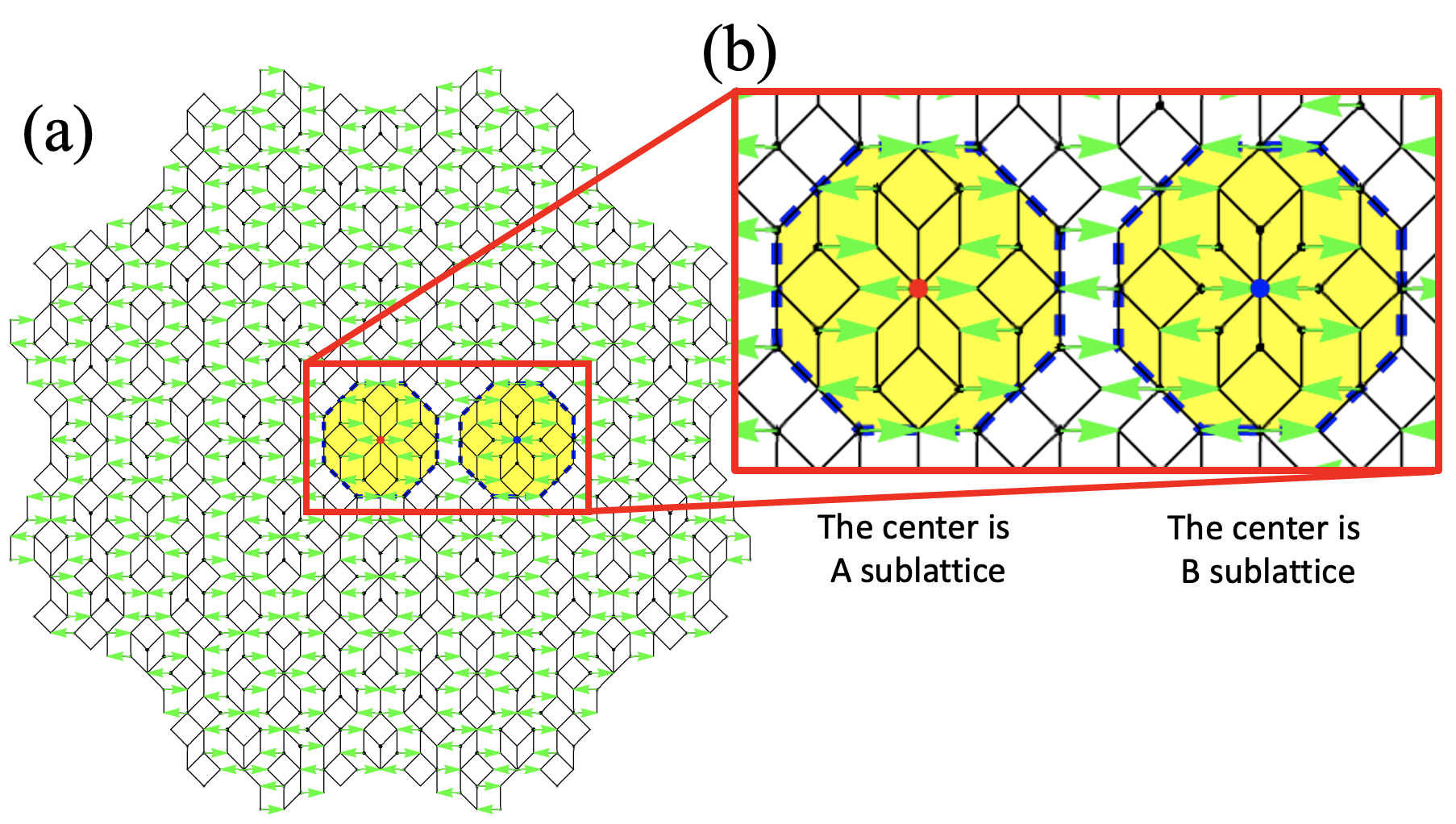}
 	\caption{(a) Schematically drawn Hamiltonian of $\pi$-flux model for $i=3$ Ammann-Beenker tiling. The hopping value from vertex $i$ to $j$ is 1, $\imath$, and $-\imath$ black lines connecting  $i$ and $j$, green arrows from $i$ to $j$, and green arrows from $j$ to $i$, respectively. (b) The area is represented by the red rectangle in (a). The yellow-shaded regions are $D^{1}$ structures.}
 	\label{fig:conf1_shiftcc}
 \end{figure}

 We discuss how the confined states in $D^{1}$ appear in $D^{i}$ for $2\leq i$. The confined state in $D^{2}$ is $\Psi^{1,\sqrt{2}}_{1}$ since $D^{2}$ includes the structure of $D^{1}$ around its center. There are 17 confined states in $D^{3}$ since $D^{3}$ includes sixteen $D^{1}$ and a single $D^{1}$ structure around its center. Figure~\ref{fig:conf1_shiftcc}~(a) illustrates the Hamiltonian of $\pi$-flux model for generation $i=3$, and the red rectangular region is magnified in Fig.~\ref{fig:conf1_shiftcc}~(b). There are two yellow shaded $D^{1}$ structures in Fig.~\ref{fig:conf1_shiftcc}~(b). When the central vertex of $D^{3}$ is A sublattice, the centers of left and right yellow-shaded $D^{1}$ structures are A and B sublattices, respectively. Inside these two yellow-shaded regions, the orientations of the arrows are dependent upon if its central vertex is A sublattice or B sublattice. By definition, the flip of the orientation of arrows corresponds to taking the complex conjugate of the Hamiltonian. The eigenstates of the complex-conjugated Hamiltonian are the complex conjugates of the eigenstates of the original Hamiltonian. Accordingly, the confined state localized around the right yellow shaded $D^{1}$ structures in Fig.~\ref{fig:conf1_shiftcc}~(b) is $\left(\Psi^{1,\sqrt{2}}_{1}\right)^{*}$. In general, the confined states configuration $c$ with energy $E$ in a $D^{i}$ structure included in $D^{j} (i<j)$ are $\left\{\Psi^{c,E}_{k}\right\}_{k}$ and $\left\{\left(\Psi^{c,E}_{k}\right)^{*}\right\}_{k}$ when the center of $D^{i}$ is on A and B sublattices, respectively.
 
 In the zero flux model on Ammann-Beenker tiling, there is a way to choose a  linear combination of confined states so that their weights are +1 or -1, at least for the smaller generations $1\leq i\leq 3$~\cite{PhysRevB.102.115125}. Although the weights are always real in the zero flux model, those in the confined states of the $\pi$-flux model can be complex. Correspondingly, we consider if there is a choice of the linear combination of confined states where the weights of confined states are one of +1, -1, $\imath$, and $-\imath$. Obviously, the components of $\Psi^{1,\sqrt{2}}_{1}$ are not in the form of +1, -1, $\imath$, and $-\imath$.
 
 The fraction of confined states in the $\pi$-flux model with energy $E$ is the same as that with energy $-E$. For example, the confined state of AB.conf1 with energy $E=-\sqrt{2}$ is $\tilde{\Psi}^{1,\sqrt{2}}_{1}$, and the corresponding fraction of confined states $p$ is the same as that of energy $E=\sqrt{2}$.

 \subsection{AB.conf2 with $E=0$}
 Next, we consider AB.conf2 [see Fig.~2(b1) in the main text] with $E=0$. For $1\leq i\leq 5$, $N_{i1}$, $N^{{\rm tot}}_{i}$, and $N^{{\rm net}}_{i}$ are shown in Table~\ref{table:conf2E1NiNet}. In this case, it is impossible to guess the general expression of the sequence of $N^{{\rm net}}_{i}$ reasonably. In the case of the zero flux  model, $N^{{\rm net}}_{i}\propto i^2$.
 On the other hand, $N^{{\rm net}}_{i}$ in the $\pi$-flux model grows faster than quadratic growth according to Table~\ref{table:conf2E1NiNet}.
 
 \begin{table}[]
 	\caption{$N_{i1}$, $N^{{\rm tot}}_{i}$, and $N^{{\rm net}}_{i}$ for $1\leq i \leq 5$ in AB.conf2 of the $\pi$-flux model on Ammann-Beenker tiling for $E=0$.}
 	\label{table:conf2E1NiNet}
 	\begin{tabular}{lll}
 		\hline\hline
 		$i$ & $N^{{\rm tot}}_{i}$ & $N^{{\rm net}}_{i}$ \\ \hline
 		1   & 1                   & 1                   \\
 		2   & 4                   & 4                   \\
 		3   & 31                  & 15                   \\
 		4   & 236                 & 68                   \\
 		5   & 1635                & 347                   \\ \hline\hline
 	\end{tabular}
 \end{table}
 
 \begin{figure}
 	\includegraphics[width=0.75\linewidth]{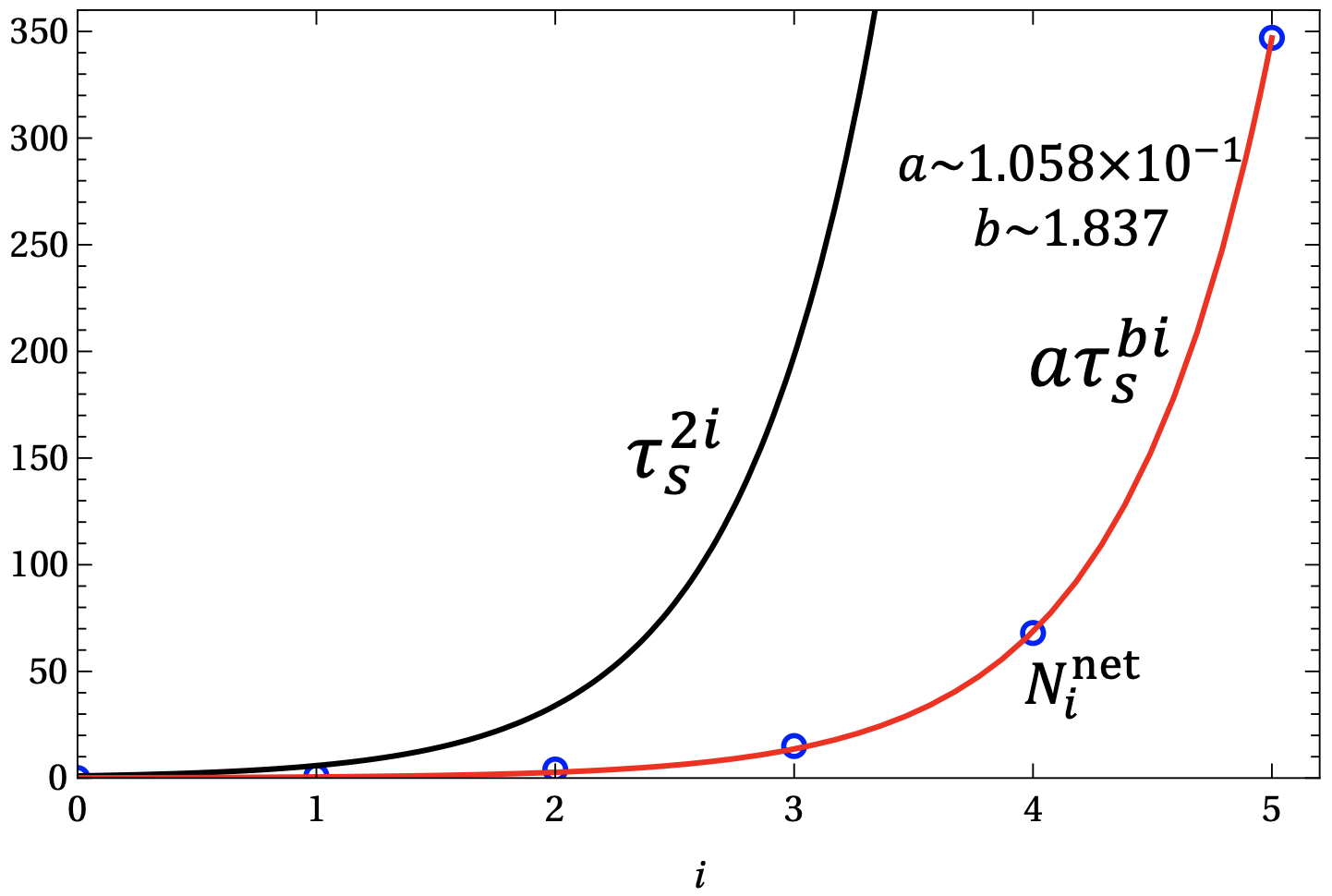}
 	\caption{$\tau_{s}^{2i}$ (black line) and $N^{{\rm net}}_{i}$ (blue circles) as a function of generation $i$. $N^{{\rm net}}_{i}$ is fitted by $a \tau_{s}^{bi}$ shown by the red line.}
 	\label{fig:NinetVStau}
 \end{figure}
 
 The sum in Eq.(\ref{eq:p_Koga}) should converge because $p\leq 1$ by definition. Because $p_{i}\propto \tau_{s}^{-2i}$, the sum in Eq.(\ref{eq:p_Koga}) does not converge if the growth speed of $N^{{\rm net}}_{i}$ is faster than $\tau_{s}^{2i}$. Accordingly, when we fit $N^{{\rm net}}_{i}$ as $a \tau_{s}^{bi}$, $b$ should be less than 2.
 We compare $N^{{\rm net}}_{i}$ with $\tau_{s}^{2i}$ in Fig.~\ref{fig:NinetVStau}. By fitting $N^{{\rm net}}_{i}$ as $a \tau_{s}^{bi}$, we obtain the fitting parameters $a\sim 1.058\times 10^{-1}$ and $b\sim 1.837$, where $b$ is less than 2.
 
\begin{figure}
 	\includegraphics[width=0.75\linewidth]{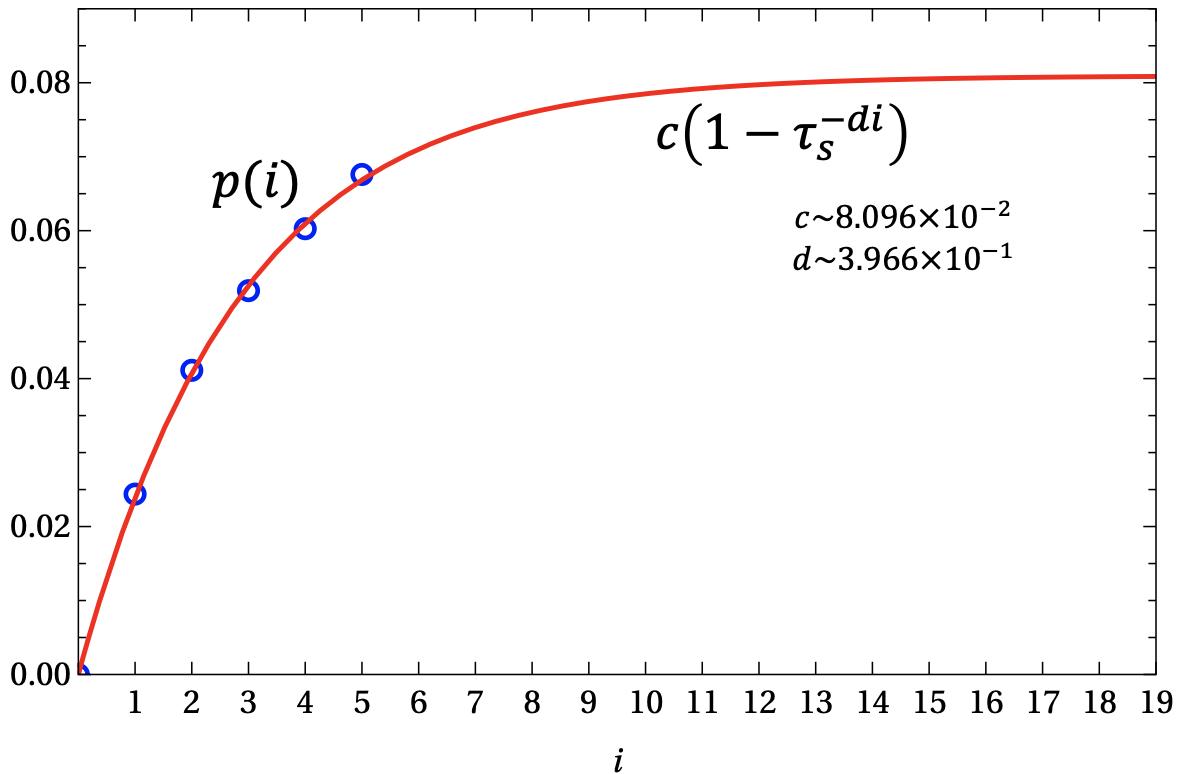}
 	\caption{The generation $i$ dependence of $p(i)$.  $p(i)$ is plotted by blue circles, while the fitted result using $c(1-\tau_{s}^{di})$ is shown by the red line.}
 	\label{fig:pofi}
 \end{figure}

 For numerically obtaining the extrapolation value of $p$ in Eq.(\ref{eq:p_Koga}), we define $p(i)$ as
 \begin{equation}
 	\label{eq:p_Koga_ofi}
 	p(i)=\sum_{j=1}^{i}p_{j}N^{{\rm{net}}}_{j}.
 \end{equation}
 In  the zero flux model, we numerically check that $p(i)$ converges faster than the fraction of $i$-th generation in the limit of $i \rightarrow \infty$. By fitting $p(i)$ as $c(1-\tau_{s}^{di})$, we obtain $c\sim 8.096 \times 10^{-2}$ and $d\sim 3.966 \times 10^{-1}$. As shown in Fig.~\ref{fig:pofi}, $p(i)$ converges to $c\sim 8.096 \times 10^{-2}$. 
 
\begin{figure}
 	\includegraphics[width=0.35\linewidth]{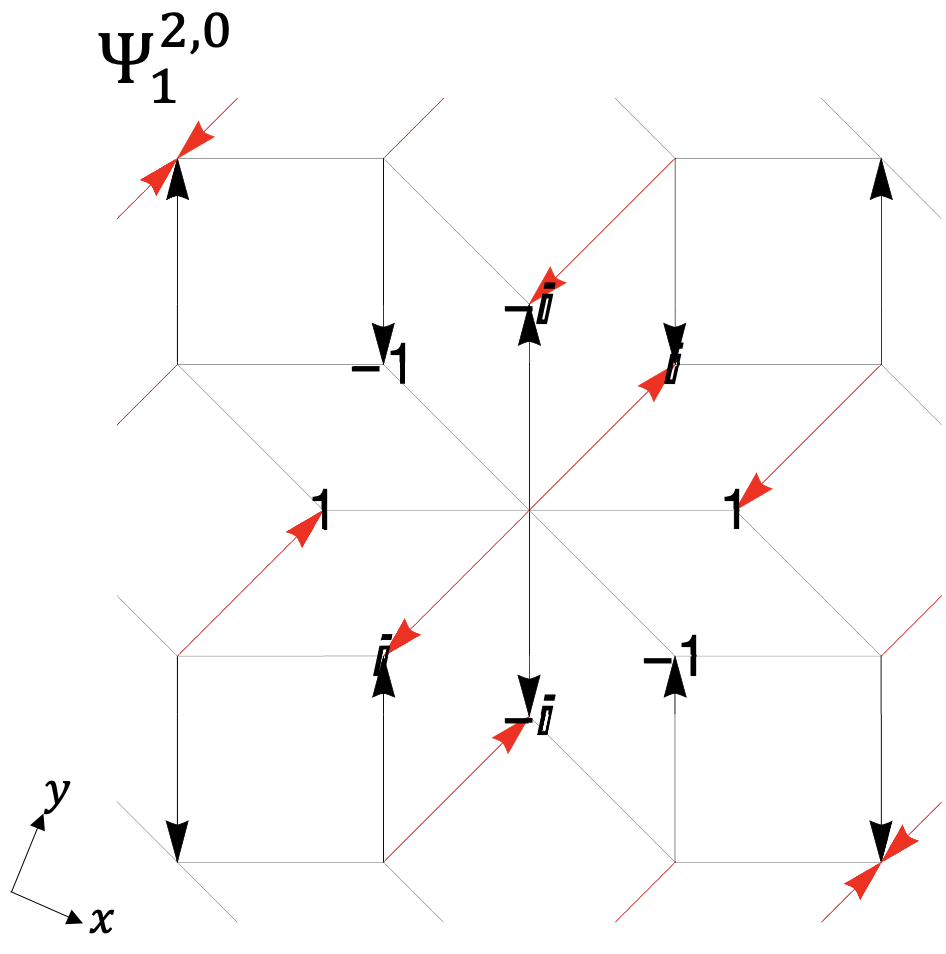}
 	\caption{The confined state in $D^{1}$ for AB.conf2 in the $\pi$-flux model on Ammann-Beenker tiling with $E=0$. Each number on the vertices represents the relative wavefunction's components} of the confined state at each vertex.
 	\label{fig:conf2E1index1}
 \end{figure}

 Next, we consider the confined state in $D^{1}$. Similar to AB.conf1, the confined states of AB.conf2 should be described by the irreducible representation of the dihedral group $D_{2}$. Here, the unit vectors for $x$ and $y$ axes are $\mbox{\boldmath $e$}_{x}=(\cos{(-\pi/8)},\sin{(-\pi/8)})$ and $\mbox{\boldmath $e$}_{y}=(\cos{(3\pi/8)},\sin{(3\pi/8)})$, respectively. The unit vector for $z$ axis, namely $\mbox{\boldmath $e$}_{z}$, is defined so that $\mbox{\boldmath $e$}_{x}, \mbox{\boldmath $e$}_{y}$, and $\mbox{\boldmath $e$}_{z}$ form the right-handed coordinate system.
 The confined state $\Psi^{2,0}_{1}$ is schematically shown in Fig.~\ref{fig:conf2E1index1}, which belongs to the irreducible representation ${\rm{B_{1}}}$.
 
\begin{figure}
 	\includegraphics[width=1\linewidth]{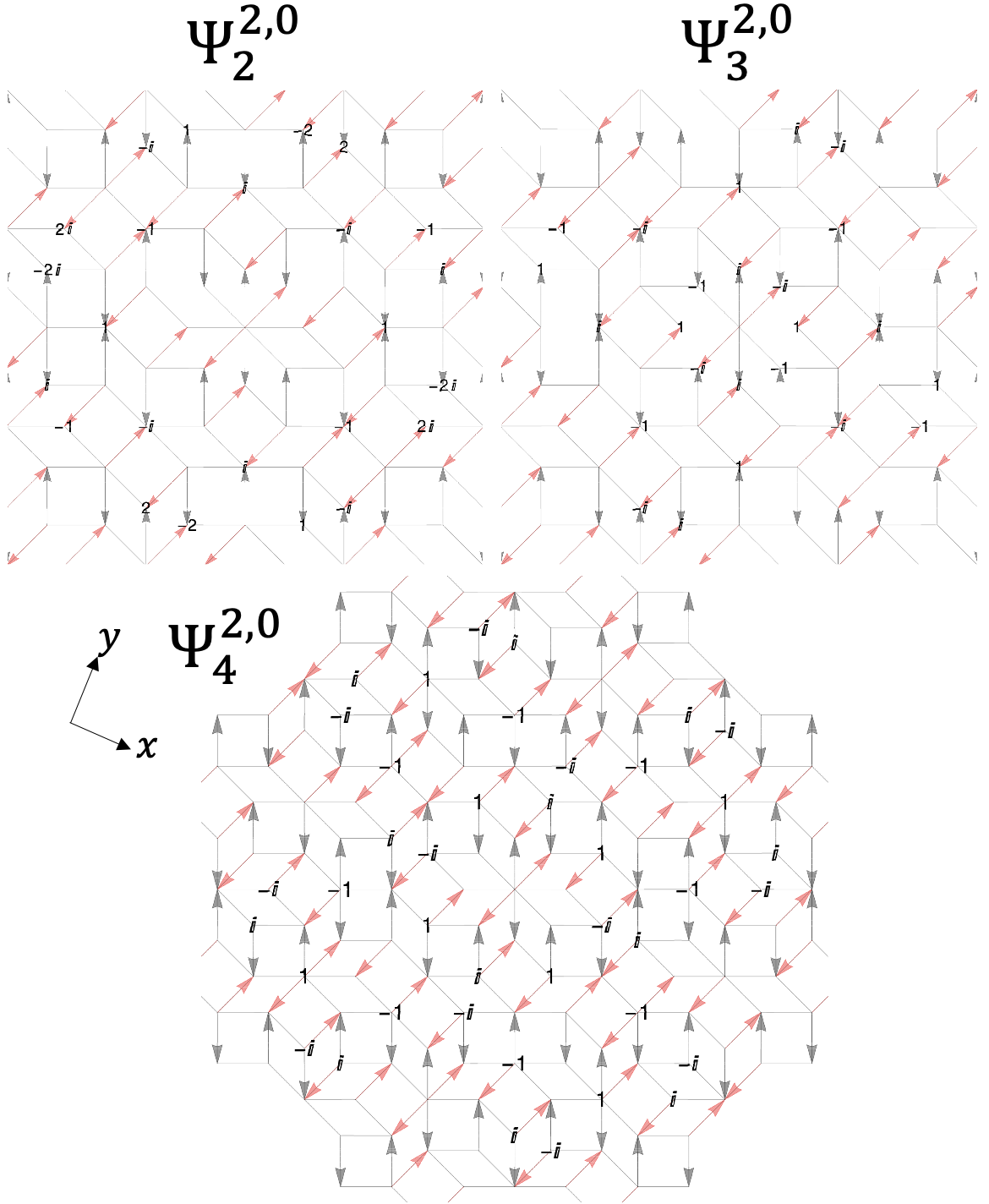}
 	\caption{The confined state in $D^{2}$ for the $\pi$-flux model of AB.conf2 with $E=0$ on Ammann-Beenker tiling. Each number on the vertices represents the amplitude of the confined state at each site. The arrows denote the hopping with nonzero Peierls phase (see the main text for definition).}
 	\label{fig:conf2E1gen2}
 \end{figure}

 The confined states in $D^{2}$ are shown in Fig.~\ref{fig:conf2E1gen2}. The confined states $\Psi^{1,\sqrt{2}}_{2}$, $\Psi^{1,\sqrt{2}}_{3}$, and $\Psi^{1,\sqrt{2}}_{4}$ belong to the irreducible representations ${\rm{B_{1}}}$, ${\rm{B_{1}}}$, and ${\rm{A}}$, respectively.
 Although $\Psi^{1,\sqrt{2}}_{1}$, $\Psi^{1,\sqrt{2}}_{2}$, and $\Psi^{1,\sqrt{2}}_{3}$ belong to the same irreducible representations ${\rm{B_{1}}}$, we have checked that these states are orthogonal each other.
 
 We note that $\{\Psi^{2,0}_{k}\}_{k}$ has amplitudes only in one of the sublattices A or B, similar to the zero flux model. The components of $\Psi^{1,\sqrt{2}}_{k}$ for $k=1,3,$ and 4 are one of +1, -1, $\imath$, and $-\imath$. But if we assume a nonzero coefficient of linear combination for $\Psi^{1,\sqrt{2}}_{2}$, there is no choice of the linear combination of confined states whose components of confined states are one of +1, -1, $\imath$, and $-\imath$.
 
 \subsection{AB.conf2 with $E=2$}
 For the AB.conf2 with $E=2$, $N_{i1}$, $N^{{\rm tot}}_{i}$, and $N^{{\rm net}}_{i}$ for $1 \leq i \leq 5$  are listed in Table~\ref{table:conf2E2NiNet}. It is possible to assume that the sequence of $N^{{\rm net}}_{i}$ is given by $N^{{\rm{net}}}_{1}=1$ and $N^{{\rm net}}_{i}=2 (2\leq i)$.
 This sequence of $N^{{\rm net}}_{i}$ implies that all of the confined states in $D^{i}$ for $3\leq i$ are the confined states in $D^{2}$, while there is a single confined state in $D^{2}$ that does not exist in $D^{1}$. We define the confined state in $D^{1}$ as $\Psi^{2,2}_{1}$ and the confined states in $D^{i} (2\leq i)$ as $\Psi^{2,2}_{1}$ and $\Psi^{2,2}_{2}$.
 From Eq.~(\ref{eq:p_Koga}), the fraction of confined states results in $p=\tau_{s}^{-4}+(\tau_{s}^{-4}-2\tau_{s}^{-5})=p^{{\rm{F}}}+(p^{{\rm{F}}}-p^{{\rm{F_{1}}}})\sim 3.449\times 10^{-2}$.
 The component $p^{{\rm{F}}}$ in $p$ is originated from $\Psi^{2,2}_{1}$ that exists in all the domains $D^{i} (1\leq i)$. On the other hand, the component $(p^{{\rm{F}}}-p^{{\rm{F_{1}}}})$ in $p$ is originated from $\Psi^{2,2}_{2}$ that exists in $D^{i} (2\leq i)$. Because $\Psi^{2,2}_{2}$ does not exist in $D^{1}$, $p^{{\rm{F_{1}}}}$ is subtracted from $p$.
 
 \begin{table}[]
 	\caption{$N_{i1}$, $N^{{\rm tot}}_{i}$, and $N^{{\rm net}}_{i}$ for $1\leq i \leq 5$ for AB.conf2 in the $\pi$-flux model on Ammann-Beenker tiling with $E=2$}.
 	\label{table:conf2E2NiNet}
 	\begin{tabular}{lll}
 		\hline\hline
 		$i$ & $N^{{\rm tot}}_{i}$ & $N^{{\rm net}}_{i}$ \\ \hline
 		1   & 1                   & 1                   \\
 		2   & 2                   & 2                   \\
 		3   & 18                  & 2                   \\
 		4   & 138                 & 2                   \\
 		5   & 874                 & 2                   \\ \hline\hline
 	\end{tabular}
 \end{table}
 
\begin{figure}
 	\includegraphics[width=1\linewidth]{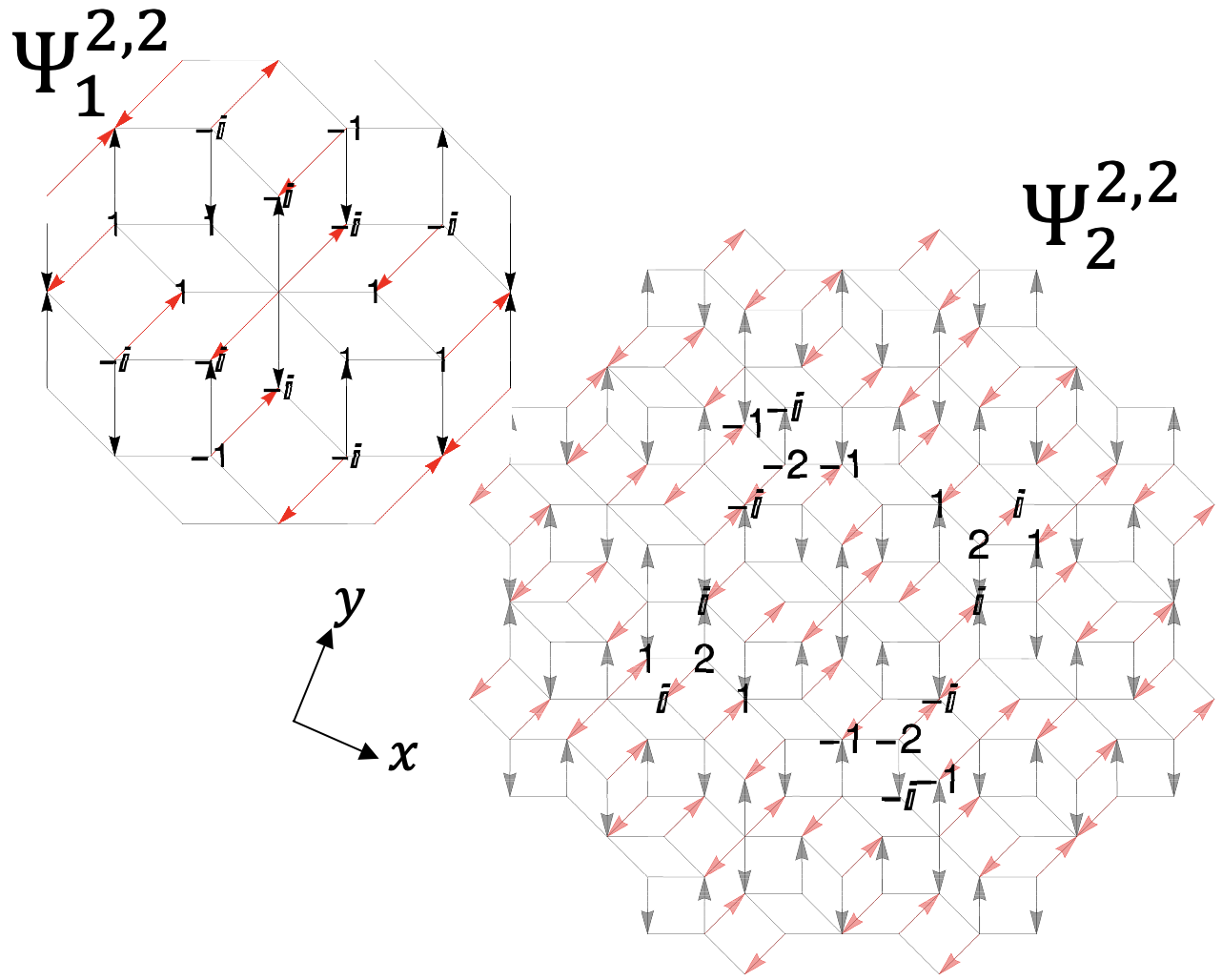}
 	\caption{Two confined states, $\Psi^{2,2}_{1}$ and $\Psi^{2,2}_{2}$, for AB.conf2 in the $\pi$-flux model on Ammann-Beenker tiling with $E=2$. Each number on the vertices represents the wavefunction's components of the confined state at each vertex. The arrows denote the hopping with nonzero Peierls phase (see the main text for definition).}
 	\label{fig:conf2E2CSall}
 \end{figure}

 In Fig.~\ref{fig:conf2E2CSall}, $\Psi^{2,2}_{1}$ and $\Psi^{2,2}_{2}$ are schematically illustrated. The irreducible representation of $\Psi^{2,2}_{1}$ and $\Psi^{2,2}_{2}$ are A and $\rm{B_{1}}$. Note that $\Psi^{2,2}_{1}$ and $\Psi^{2,2}_{2}$ have amplitudes in both sublattices A and B.
 The components of $\Psi^{2,2}_{1}$ ($\Psi^{2,2}_{2}$) are (not) in the form of +1, -1, $\imath$, and $-\imath$.
 
 \subsection{AB.conf3 with $E=0$}
 For the AB.conf3 [see Fig.~2(c1) in the main text] with $E=0$, $N_{i1}$, $N^{{\rm tot}}_{i}$, and $N^{{\rm net}}_{i}$ for $1 \leq i \leq 5$  are listed in Table~\ref{table:conf3E1NiNet}. Judging from $N^{{\rm net}}_{i}$, we speculate $N^{{\rm net}}_{i}=7$ more than $i=5$.
 We define the confined state in $D^{1}$ as $\Psi^{3,0}_{1}$, the confined states in $D^{2}$ as $\left\{\Psi^{3,0}_{k}\right\}_{k} (1\leq k \leq 5)$, and the confined states in $D^{3}$ as $\left\{\Psi^{3,0}_{k}\right\}_{k} (1\leq k \leq 7)$.
 The fraction of confined states is obtained as $p=\tau_{s}^{-4}+4(\tau_{s}^{-4}-2\tau_{s}^{-5})+2(\tau_{s}^{-4}-2\tau_{s}^{-5}-2\tau_{s}^{-7})=p^{{\rm{F}}}+4(p^{{\rm{F}}}-p^{{\rm{F_{1}}}})+2(p^{{\rm{F}}}-p^{{\rm{F_{1}}}}-p^{{\rm{F_{2}}}})\sim 5.137\times 10^{-2}$.
 The component $p^{{\rm{F}}}$ in $p$ is the contribution of $\Psi^{3,0}_{1}$ that exist in all the domains $D^{i} (1\leq i)$. The component $4(p^{{\rm{F}}}-p^{{\rm{F_{1}}}})$ in $p$ is the contribution of four confined states that exist in $D^{i} (2\leq i)$, namely, $\Psi^{3,0}_{2}$, $\Psi^{3,0}_{3}$, $\Psi^{3,0}_{4}$, and $\Psi^{3,0}_{5}$.
 Also, the component $2(p^{{\rm{F}}}-p^{{\rm{F_{1}}}}-p^{{\rm{F_{2}}}})$ in $p$ is the contribution of two confined states that exist in $D^{i} (3\leq i)$, namely, $\Psi^{3,0}_{6}$ and $\Psi^{3,0}_{7}$. Because both $\Psi^{3,0}_{6}$ and $\Psi^{3,0}_{7}$ do exist neither in $D^{1}$ nor $D^{1}$, $2p^{{\rm{F_{1}}}}$ and $2p^{{\rm{F_{2}}}}$ are subtracted from $p$.
 
 \begin{table}[]
 	\caption{$N_{i1}$, $N^{{\rm tot}}_{i}$, and $N^{{\rm net}}_{i}$ for $1\leq i \leq 5$ for AB.conf3 in the $\pi$-flux model on Ammann-Beenker tiling with $E=0$.}
 	\label{table:conf3E1NiNet}
 	\begin{tabular}{lll}
 		\hline\hline
 		$i$ & $N^{{\rm tot}}_{i}$ & $N^{{\rm net}}_{i}$ \\ \hline
 		1   & 1                   & 1                   \\
 		2   & 5                   & 5                   \\
 		3   & 23                  & 7                   \\
 		4   & 191                 & 7                   \\
 		5   & 1271                 & 7                   \\ \hline\hline
 	\end{tabular}
 \end{table}
 
 The confined states in AB.conf3 belong to the irreducible representation of the dihedral group $D_{4}$. Table~\ref{table:D4char} is a character table for the dihedral group $D_{4}$.
 
 \begin{table}[]
 	\caption{A character table for the dihedral group $D_{4}$, where $E$ is an identity operator, $C_{4}(z)$ is a rotation operator of $\pi/2$ about $z$-axis, $C_{2}(z)$ is a rotation operator of $\pi$ about $z$-axis, $C'_{2}$ and $C''_{2}$ are umklappung operators, namely, $C'_{2}$ and $C''_{2}$ are rotation operators of $\pi$ about $x$-axis and the axis whose unit vector is $(\mbox{\boldmath $e$}_{x}+\mbox{\boldmath $e$}_{y})/\sqrt{2}$, respectively.
 		Here, the unit vectors for $x$ and $y$ axes are $\mbox{\boldmath $e$}_{x}=(1,0)$ and $\mbox{\boldmath $e$}_{y}=(0,1)$, respectively. The unit vector for $z$ axis, namely $\mbox{\boldmath $e$}_{z}$, is defined so that $\mbox{\boldmath $e$}_{x}, \mbox{\boldmath $e$}_{y}$, and $\mbox{\boldmath $e$}_{z}$ form the right-handed coordinate system.}
 	\label{table:D4char}
 	
 	\begin{tabular}{llllll}
 		\hline\hline
 		& $E$ & $2C_{4}(z)$ & $C_{2}(z)$ & $2C'_{2}$ & $2C''_{2}$ \\ \hline
 		$\rm{A_{1}}$ & +1  & +1          & +1         & +1        & +1         \\
 		$\rm{A_{2}}$ & +1  & +1          & -1         & -1        & -1         \\
 		$\rm{B_{1}}$ & +1  & -1          & +1         & +1        & -1         \\
 		$\rm{B_{2}}$ & +1  & -1          & +1         & -1        & +1         \\
 		E            & +2  & 0           & -2         & 0         & 0          \\ \hline\hline
 	\end{tabular}
 \end{table}
 
\begin{figure}
 	\includegraphics[width=1\linewidth]{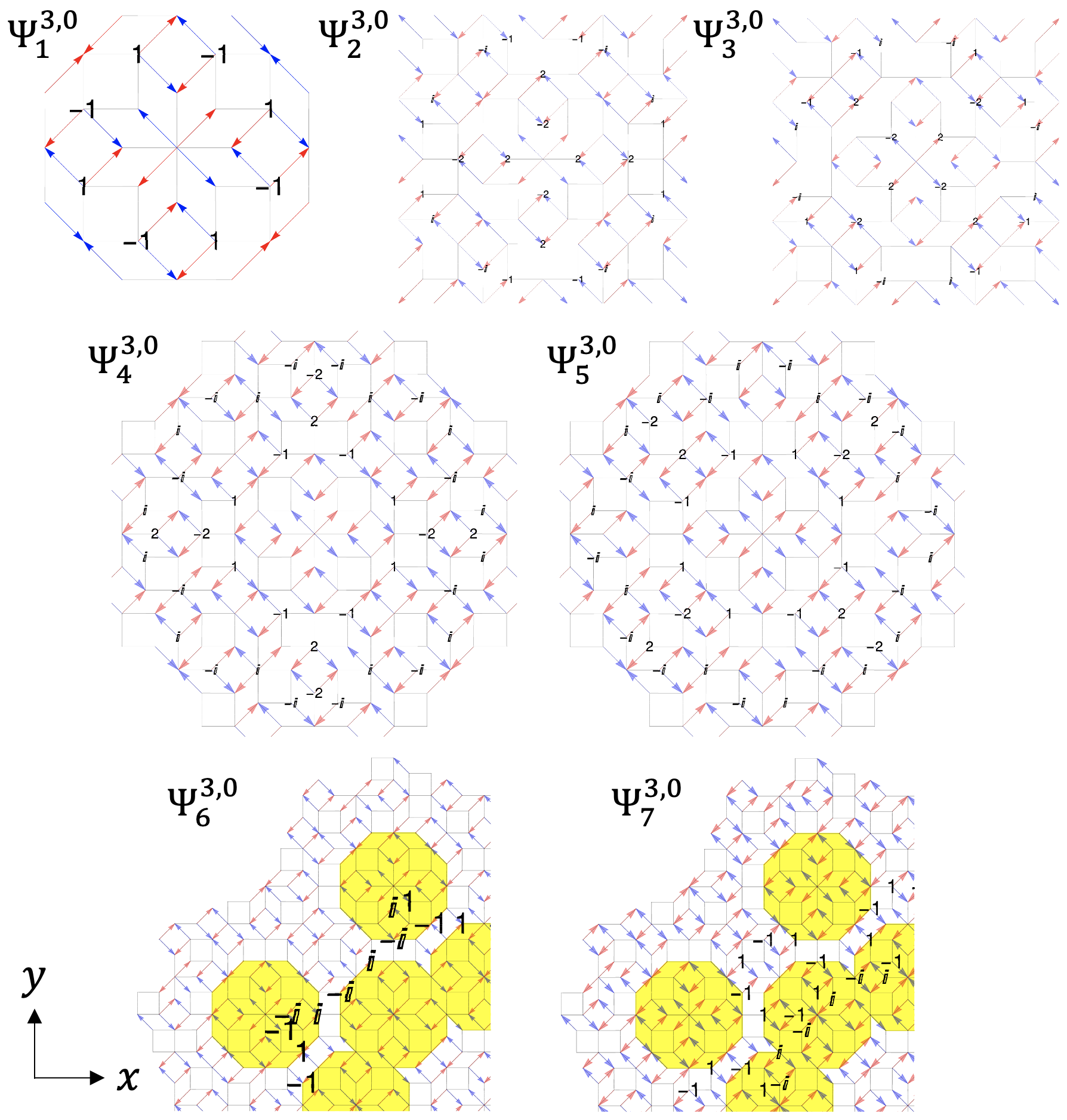}
 	\caption{Seven confined states, $\left\{\Psi^{3,0}_{k}\right\}_{k} (1\leq k \leq 7)$, for AB.conf3 in the $\pi$-flux model on Ammann-Beenker tiling with $E=0$.  Utilizing the even or odd mirror symmetry of confined states, one-fourth of the entire tiling of $\Psi^{3,0}_{6}$ and $\Psi^{3,0}_{7}$ are shown. Each number on the vertices represents the amplitude of the confined state at each vertex.
  The arrows denote the hopping with nonzero Peierls phase (see the main text for definition).
 	}
 	\label{fig:conf3E1all_zoom}
 \end{figure}

 In Fig.~\ref{fig:conf3E1all_zoom}, $\left\{\Psi^{3,0}_{k}\right\}_{k} (1\leq k \leq 7)$ are schematically shown. The irreducible representations of $\Psi^{3,0}_{k}$ for $k=1,2,3,4,5,6,$ and 7 are ${\rm A_{2}}$, ${\rm B_{1}}$, ${\rm B_{2}}$, ${\rm B_{1}}$, ${\rm B_{2}}$, ${\rm A_{2}}$, and ${\rm A_{2}}$, respectively.  Note that $\left\{\Psi^{3,0}_{k}\right\}_{k} (1\leq k \leq 7)$ have amplitudes in one of the sublattices A or B only, similar to the zero flux model on Ammann-Beenker tiling.
 There is a way to choose the linear combination of confined states so that their wavefunction's components are one of +1, -1, $\imath$, and $-\imath$. In detail, $\Psi^{3,0}_{1}$, $(\Psi^{3,0}_{2}+\imath \Psi^{3,0}_{3})/2$, $(\Psi^{3,0}_{2}-\imath \Psi^{3,0}_{3})/2$, $(\Psi^{3,0}_{4}+ \Psi^{3,0}_{5})/2$, $(\Psi^{3,0}_{4}-\imath \Psi^{3,0}_{5})/2$, $\Psi^{3,0}_{6}$, and $\Psi^{3,0}_{7}$ are in the form of +1, -1, $\imath$, and $-\imath$.
 It is interesting that since $\Psi^{3,0}_{1}\in \mathbb{Z}$ satisfies $\left(\Psi^{3,0}_{1}\right)^{*}=\Psi^{3,0}_{1}$, $\Psi^{3,0}_{1}$ is a confined state regardless of the sublattice type of the central vertex in $D^{1}$.
 
 \subsection{AB.conf3 with $E=2\sqrt{2}$}
 For the AB.conf3 with $E=2\sqrt{2}$, $N_{i1}$, $N^{{\rm tot}}_{i}$, and $N^{{\rm net}}_{i}$ for $1 \leq i \leq 5$  are listed in Table~\ref{table:conf3E2NiNet}. Similar to AB.con1 with  $E=\sqrt{2}$ $N^{{\rm net}}_{i}$ should be 1 for $1\leq i$.
 Accordingly, the fraction of confined states results in $p=\tau_{s}^{-4}=p^{{\rm{F}}}\sim 2.944\times 10^{-2}$. There is a single type of the confined state $\Psi^{3,2\sqrt{2}}_{1}$ that is localized around the central vertex of $D^{1}$.
 
 \begin{table}[]
 	\caption{$N_{i1}$, $N^{{\rm tot}}_{i}$, and $N^{{\rm net}}_{i}$ for $1\leq i \leq 5$ in AB.conf3 of $\pi$-flux model on Ammann-Beenker tiling with $E=2\sqrt{2}$.}
 	\label{table:conf3E2NiNet}
 	\begin{tabular}{lll}
 		\hline\hline
 		$i$ & $N^{{\rm tot}}_{i}$ & $N^{{\rm net}}_{i}$ \\ \hline
 		1   & 1                   & 1                   \\
 		2   & 1                   & 1                   \\
 		3   & 17                  & 1                   \\
 		4   & 121                 & 1                   \\
 		5   & 753                 & 1                   \\ \hline\hline
 	\end{tabular}
 \end{table}
 
\begin{figure}
 	\includegraphics[width=0.35\linewidth]{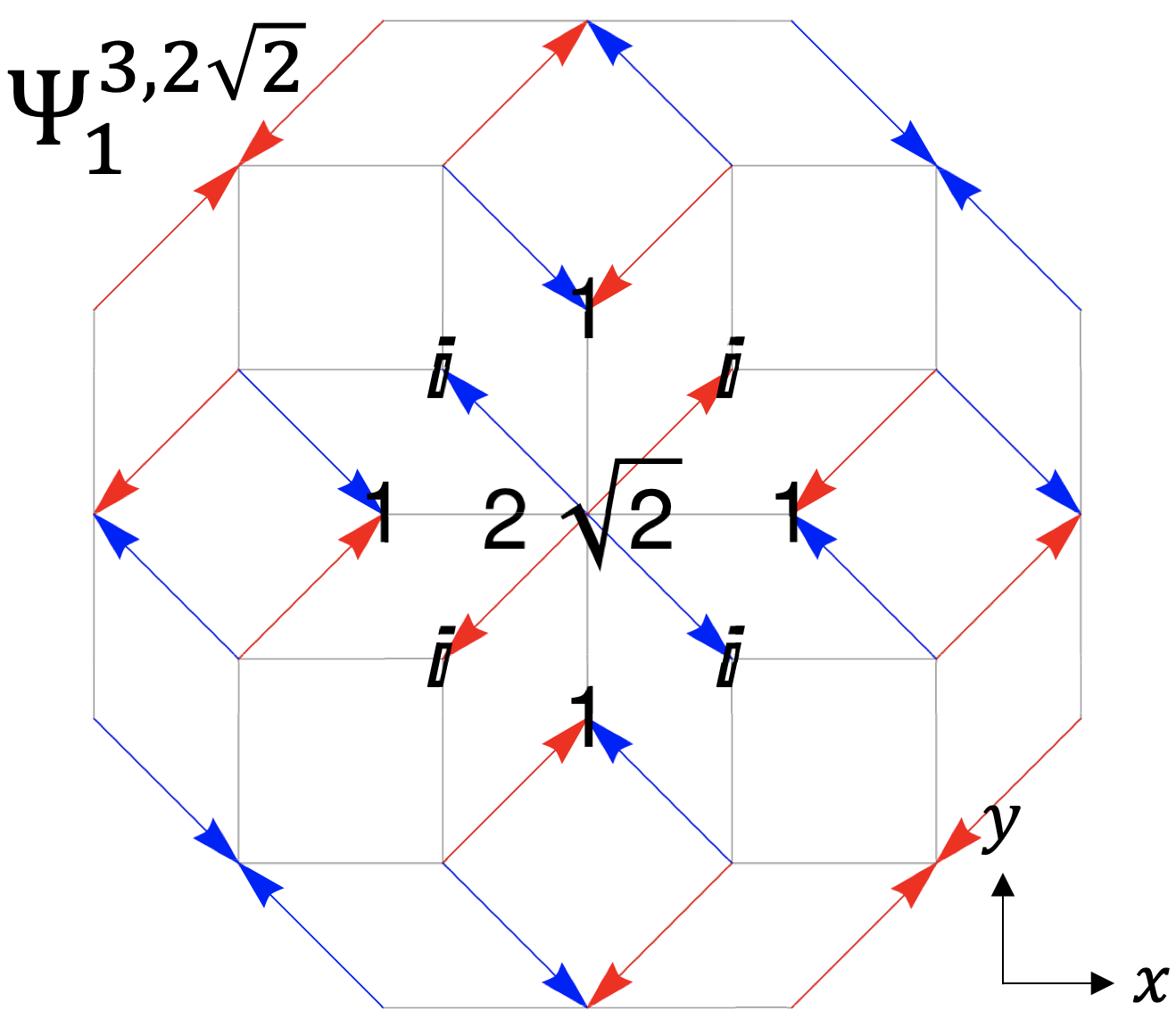}
 	\caption{The confined state, $\Psi^{3,2\sqrt{2}}_{1}$, for AB.conf3 in the $\pi$-flux model on Ammann-Beenker tiling with $E=2\sqrt{2}$. Each number on the vertices represents the amplitude of the confined state at each vertex.
  The arrows denote the hopping with nonzero Peierls phase (see the main text for definition).
 	}
 	\label{fig:conf3E2}
 \end{figure}

 The confined state $\Psi^{3,2\sqrt{2}}_{1}$ is schematically shown in Fig.~\ref{fig:conf3E2}. Its irreducible representation is ${\rm A_{1}}$. Note that $\Psi^{3,2\sqrt{2}}_{1}$ has amplitudes in both sublattices A and B. Obviously, the components of $\Psi^{3,2\sqrt{2}}_{1}$ are not in the form of +1, -1, $\imath$, and $-\imath$.
 
 \subsection{Summary for all the configurations}
\begin{figure}
 	\includegraphics[width=1\linewidth]{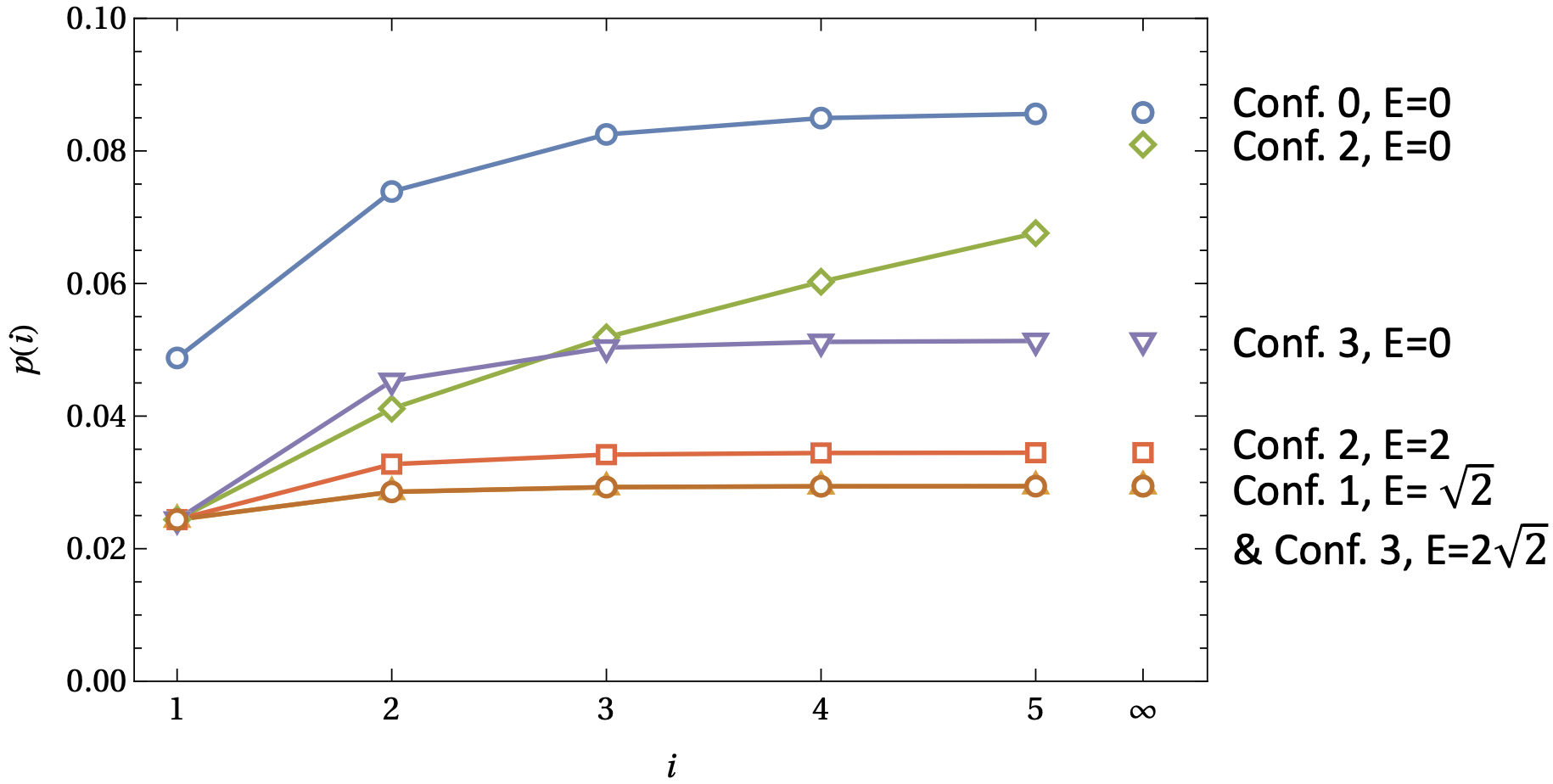}
 	\caption{The generation $i$ dependence of $p(i)$ for the $\pi$-flux model on Ammann-Beenker tiling. ``Conf.$c$, $E=e$'' corresponds to the configuration $c$ with energy $e$, where configuration 0 is the zero flux  model. The converged value of $\lim_{i \to \infty}p(i)$, which is obtained by calculating $p$ in Eq.(\ref{eq:p_Koga}) except for the AB.conf2 with $E=0$, is plotted at $i=\infty$. The converged value of $\lim_{i \to \infty}p(i)$ for the AB.conf2 with $E=0$ is obtained by the fitting in Fig.~\ref{fig:pofi}.}
 	\label{fig:pi_allConf}
 \end{figure}

 We summarize the generation $i$ dependence of $p(i)$ for the $\pi$-flux model in Fig.~\ref{fig:pi_allConf}, where Conf.0 corresponds to the zero flux model. We plot $p=\lim_{i \to \infty}p(i)$ at $i=\infty$. It is found that $p$ for any configuration in the $\pi$-flux model is smaller than that of the zero flux model. The behavior of convergence in $p(i)$ is almost the same for all the configurations and $p(5) \sim p$, except for AB.conf2 with $E=0$.
 
 Table.~\ref{table:confProp} summarizes the properties of confined states of the zero flux  model and the $\pi$-flux model on Ammann-Beenker tiling. The amplitudes of the confined states for zero (nonzero) energy are (not) on one of the sublattices A or B.  Whether the energy of the confined states is zero or not is irrelevant to whether or not there is a way to choose confined states so that they are one of +1,-1,$\imath$, and $-\imath$ only.
 
 \begin{table*}[]
 	\caption{A summary of the properties of confined states of the zero flux model and the $\pi$-flux model on Ammann-Beenker tiling. ``Conf.'' shows the configuration, ``Dihedral group'' shows the dihedral group by which the confined states are described, $E$ shows the energy of confined states, ``A or B only?'' shows if the amplitudes of the confined states are only on one of the sublattices A or B, ``$\pm 1, \pm \imath 1$ only?'' shows if there is a way to choose confined states so that they are one of +1,-1,$\imath$, and $-\imath$ only, \and $p$ is the converged value of the fraction of confined states calculated by Eq.(\ref{eq:p_Koga}) except for the AB.conf2 with $E=0$. Y and N correspond to yes and no, respectively.
 	}
 	\label{table:confProp}
 	
 	\begin{tabular}{|l|llcccl|}
 		\hline
 		Conf.                   & 0              & 1            & \multicolumn{2}{c}{2}                        & \multicolumn{2}{c|}{3}                                                                               \\
 		Dihedral group          & $D_{8}$        & $D_{2}$      & \multicolumn{2}{c}{$D_{2}$}                  & \multicolumn{2}{c|}{$D_{4}$}                                                                         \\
 		$E$                     & 0              & $\sqrt{2}$   & 0 & 2                                        & 0                                                                                     & $2\sqrt{2}$  \\
 		A or B only?            & Y              & N            & Y & N                                        & Y                                                                                     & N            \\
 		$\pm 1,\pm\imath$ only? & Y              & N            & N & N                                        & Y                                                                                     & N            \\
 		$p$                     & $1/2 \tau^{2}$ & $p^{\rm{F}}$ &  ? & \begin{tabular}{c}
 			$p^{\rm{F}}$\\
 			$+(p^{\rm{F}}-p^{\rm{F_{1}}})$
 		\end{tabular} & \begin{tabular}{c}
 			$p^{\rm{F}}$\\
 			$+4(p^{\rm{F}}-p^{\rm{F_{1}}})$\\
 			$+2(p^{\rm{F}}-p^{\rm{F_{1}}}-p^{\rm{F_{2}}})$
 		\end{tabular}
 		& $p^{\rm{F}}$ \\ \hline
 	\end{tabular}
 \end{table*}

	\bibliography{MHreference}
\end{document}